\documentclass[journal=jacsat,manuscript=article]{achemso}

\usepackage[version=3]{mhchem} 
\usepackage[utf8]{inputenc}
\usepackage[T1]{fontenc}
\usepackage{soul}
\usepackage{bm}
\usepackage{threeparttable}
\usepackage{color}
\usepackage{booktabs}
\usepackage{lscape}
\usepackage{multirow}

\newcommand{\cm}{cm$^{-1}$}

\author{Apurba Nandi}
\affiliation{Department of Physics and Materials Science, University of Luxembourg, L-1511, Luxembourg City, Luxembourg.}
\email{apurba.nandi@uni.lu}
\author{Riccardo Conte}
\email{riccardo.conte1@unimi.it}
\affiliation{Dipartimento di Chimica, Universit\`{a} degli Studi di Milano, via Golgi 19, 20133 Milano, Italy}
\author{Priyanka Pandey}
\affiliation{Department of Chemistry and Cherry L. Emerson Center for Scientific Computation, Emory University, Atlanta, Georgia 30322, USA.}
\author{Paul L. Houston}
\affiliation{Department of Chemistry and Chemical Biology, Cornell University, Ithaca, New York
14853, USA and Department of Chemistry and Biochemistry, Georgia Institute of
Technology, Atlanta, Georgia 30332, USA}
\author{Chen Qu}
\affiliation{Independent Researcher, Toronto, Ontario M9B0E3, Canada}

\author{Qi Yu}
\affiliation{Department of Chemistry, Fudan University, Shanghai, 200438, P. R. China }
\author{Joel M. Bowman}
\email{jmbowma@emory.edu}
\affiliation{Department of Chemistry and Cherry L. Emerson Center for Scientific Computation, Emory University, Atlanta, Georgia 30322, USA.}

\title[]{The quantum nature of ubiquitous vibrational features revealed for ethylene glycol}

\abbreviations{PES,PIP}

\begin{document}


\newpage
\begin{abstract}
Vibrational properties of molecules are of widespread interest and importance in chemistry and biochemistry. The reliability of widely employed approximate computational methods is questioned here against the complex experimental spectrum of ethylene glycol. Comparisons between quantum vibrational self-consistent field and virtual-state configuration interaction (VSCF/VCI), adiabatically switched semiclassical initial value representation (AS SCIVR), and thermostatted ring polymer molecular dynamics (TRPMD) calculations are made using a full-dimensional machine-learned potential energy surface. Calculations are done for five low-lying conformers and compared with the experiment, with a focus on the high-frequency, OH-stretches, and CH-stretches, part of the spectrum. Fermi resonances are found in the analysis of VSCF/VCI eigenstates belonging to the CH-stretching band. Results of comparable accuracy, quality, and level of detail are obtained by means of AS SCIVR.  The current VSCF/VCI and AS-SCIVR power spectra largely close the gaps between the experiment and  TRPMD and classical MD calculations.  Analysis of these results provide guidance on what level of accuracy to expect from TRPMD and classical MD calculations of the vibrational spectra for ubiquitous CH and OH-stretches bands. This work shows that even general vibrational features require a proper quantum treatment usually not achievable by the most popular theoretical approaches.
\end{abstract}

\section{Introduction}
\begin{figure}
    \centering
    \includegraphics[width=0.9\linewidth]{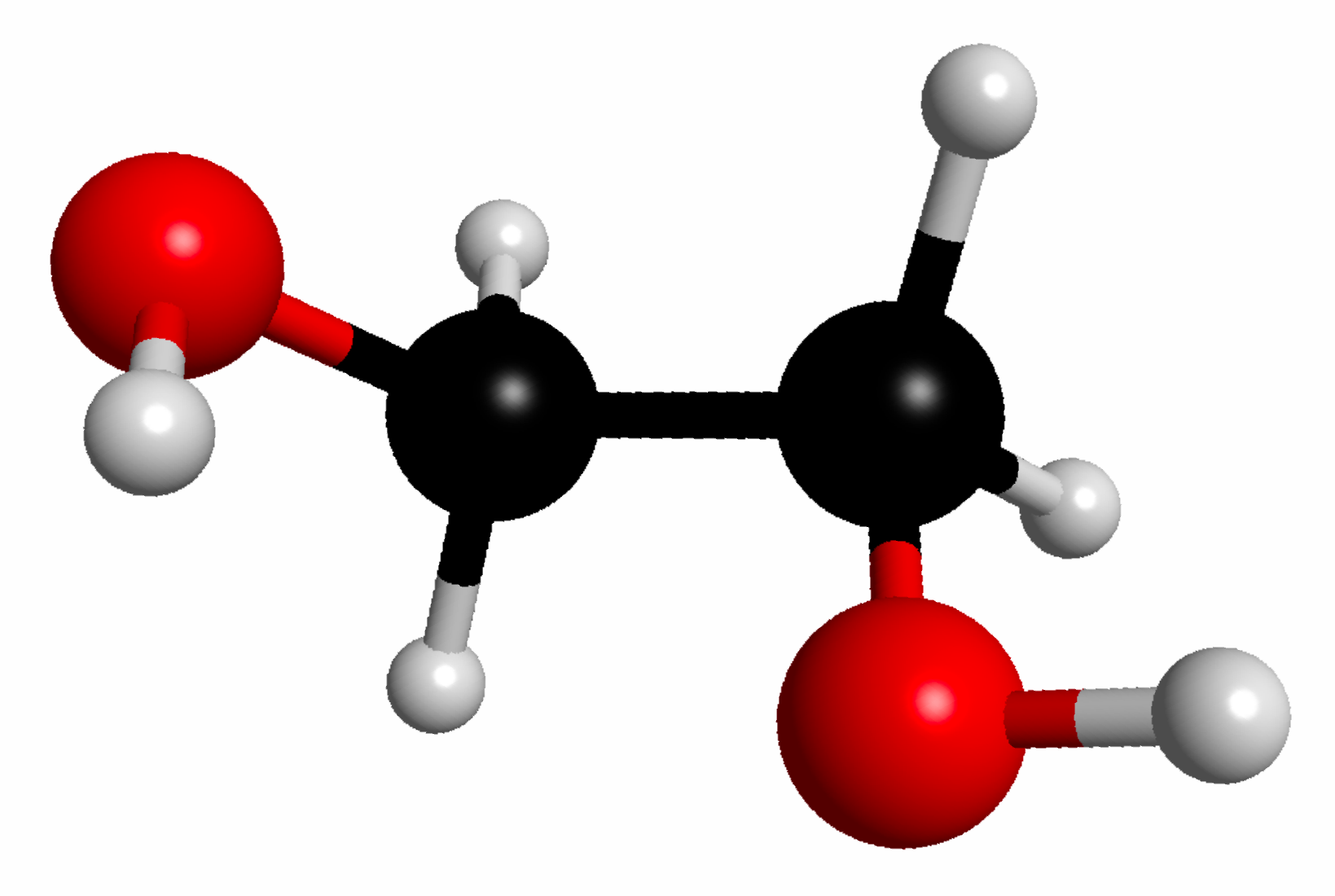}
    \caption{Global minimum structure of ethylene glycol.  }
    \label{fig:structure}
\end{figure}

The importance of intramolecular hydrogen bonding on the conformation of biomolecules is well-established.  As discussed in the literature, ethylene glycol, depicted in Fig. \ref{fig:structure}, has attracted the attention of experimentalists as being a small molecule with two vicinal hydroxyl groups where this can be studied in detail, at least in principle.  In practice, the numerous low-lying conformers of the ethylene glycol complicate the analysis of experiments, as emphasized in the recent paper by Das et al.\cite{das2015}  That paper focuses on the signature of this bonding in the IR spectrum under conditions of low concentration in the gas phase at 303, 313, and  323 K. In addition, an analysis of the thermal contribution of ten low-lying conformers to the IR spectrum was made using the double-harmonic approximation based on DFT calculations (B3LYP/aug-cc-pVDZ).

This molecule has also been studied by several theoretical chemists with great interest due to controversy on the existence of intramolecular hydrogen bonding and its complex torsional landscape via three torsional degrees of freedom (the \ce{OCCO} and two \ce{HOCC} dihedrals).

Recently, Arandhara and Ramesh reported an interesting study of quantum effects in the temperature-dependent structure of ethylene glycol. They used path integral and classical molecular dynamics as well as classical and thermostated ring-polymer molecular dynamics (TRMPD)\cite{trpmd} of the vibrational power spectrum of ethylene glycol, using a new full-dimensional potential energy surface (PES).\cite{Sai_2024} The PES is a fit to 18,772 MP2/aug-cc-pVTZ energies concentrated in the region of a reduced dimensionality space determined by minimizing the energy with respect to three dihedral angles. The full PES is given by the sum $V_{rs}$ + $V_b$, where $V_{rs}$ is the three-degree-freedom minimized potential referred as `reaction surface' potential and $V_b$ is a power-fit to energies displaced from the minimized surface. We omit the details of this elaborate representation and refer the interested reader to their paper and supporting information for details.\cite{Sai_2024} We do use this PES for limited calculations, as described below. 

The classical and TRPMD vibrational spectra reported using this PES are of particular interest to us, as these relate to experimental IR spectra and also motivate the work we present here. These spectra were compared for the signature OH and CH-stretch bands at 300 K, where several conformers contribute significantly to these spectra.  In both the classical and TRPMD approaches the power spectrum is obtained from the Fourier transform of the velocity autocorrelation function, either at fixed total energy in the case of microcanonical classical MD or fixed temperature in the case of TRMPD.\cite{trpmd}  TRPMD uses ring-polymer molecular dynamics\cite{rpmd} coupled to a thermostat\cite{Rossi2018} as a means to obtain quantum thermal effects, mainly zero-point energy (ZPE) effects. There is no explicit quantization of excited vibrational motion in TRPMD, and if a single bead is used then TRPMD becomes canonical classical MD. As we review below, both calculated spectra are upshifted from the experiment, with the classical one more so than the TRPMD one.  Of course, this may be due to errors inherent in both approaches or the PES or both.  Evidence for the former was recently presented by Qi and Bowman for \ce{H7O3+} and \ce{H9O4+}.\cite{yutests} In that work the IR spectrum was calculated using these methods and vibrational self-consistent field/virtual-state configuration interaction (VSCF/VCI)\cite{Bowman1978, vscf86, VCI} calculations, all using accurate potential and dipole moments surfaces.  Excellent agreement with the experiment was seen with the VSCF/VCI calculations.  These closed a large gap between the experiment and classical and also the smaller gap with TRPMD, both with respect to band positions and widths. 

Here we report a new fit to these electronic energies and gradients using our permutationally invariant polynomial (PIP) approach.\cite{Braams2009, ARPC2018} The new PES is used in VSCF/VCI and adiabatically switched semiclassical initial value representation (AS SCIVR)\cite{Conte_Ceotto_ASSCIVR_2019, Botti_Conte_ASotf_2022} calculations of the power spectrum. Calculations are done for five low-lying conformers and compared with the experiment, with a focus on the high-frequency, OH, and CH-stretches, part of the spectrum. Fermi resonances are found in the analysis of VSCF/VCI eigenstates belonging to the CH-stretching band. Results of comparable accuracy, quality, and level of detail are obtained by means of AS SCIVR. Comparisons are also made with classical and TRPMD calculations of the vibrational power spectrum by Mrinal and Ramesh, using their fit to these electronic energies.  The current VSCF/VCI and AS-SCIVR power spectra largely close the gap between the experiment and these previous classical and TRPMD calculations. Discussion of these comparisons sheds additional light on the limitations of these methods. 

The paper is organized as follows.  A brief review of the theoretical methods is given followed by computational details.  Following that results and discussion are given, and we conclude with a summary and conclusions.

\section{Theory and Computational Details}
\subsection{Linear Regression with Permutationally Invariant Polynomials}

Here we employ the well-established permutationally invariant polynomial (PIP) approach\cite{Braams2009,msachen, Houston2023PIPsoftware} to fit the full-dimensional PES of ethylene glycol.
The theory of this PIP approach has been presented in several review articles\cite{Braams2009, Bowman2010, Xie10, bowman11, ARPC2018} and therefore we are not presenting it in great detail.
In terms of a PIP basis, the potential energy, $V$, can be written in compact form as
\begin{equation}
\label{eq:2}
V(\bm{x})= \sum_{i=1}^{n_p} c_i p_i(\bm{x}),
\end{equation}
where $c_i$ are linear coefficients, $p_i$ are PIPs, $n_p$ is the total number of polynomials for a given maximum polynomial order, and $\bm{x}$ are Morse variables.  For example, $x_{\alpha \beta}$ is given by $\exp(-r_{\alpha \beta}/\lambda)$, where $r_{\alpha \beta}$ is the internuclear distance between atoms $\alpha$ and $\beta$. The range (hyper)parameter, $\lambda$, was chosen as 2 bohr. 

Optimal parameters such as the coefficients ($c$) are obtained by minimizing the loss function.
\begin{equation}
\label{eq:3}
L(c) := \sum_{X} (w_X^E|E(c; X) - E_{QM}(X)|^2 + w_X^F|F(c; X) - F_{QM}(X)|^2)
\end{equation}
Where $E_{QM}$ and $F_{QM}$ are the energies and corresponding forces in the training dataset, obtained from direct electronic structure calculations. The sum is taken over all configurations in the training dataset, and $w_X^E$ and $w_X^F$ are the weights specifying the relative importance of energies and forces. Here we use equal weights for both energy and forces ($w_X^E$ = $w_X^F$ = 1). As energy
and force are both linear in the free parameters, the loss can be written in a linear least-squares form
\begin{equation}
\label{eq:4}
L(c) := || \xi c - t || ^2
\end{equation}
where the vector $t$ contains the direct QM energy and force
observations and the design matrix $\xi$ contains the values of the PIP basis and
the negative gradients of the PIP basis evaluated at the training geometries. The number of rows of $\xi$ equal to the total number of observations (energies and force components) in the training
dataset and the number of columns equal to the total number of
basis functions. More often this linear regression problem can be regularized by modifying the loss function as 
\begin{equation}
\label{eq:5}
L(c) := || \xi c - t || ^2 + \eta||\Gamma c||^2,
\end{equation}
where $\Gamma$ is the identity matrix. However, we did not regularize the loss $L(c)$ function as the number of rows of the $\xi$ matrix is much larger than the number of unknown coefficients $c$.

In order to develop the PES a total of 18,772 MP2/aug-cc-pVTZ energies and the corresponding forces (a total data size of 581,932) are employed. This dataset was generated by Arandhara and  Ramesh and we have taken from their recently reported article.\cite{Sai_2024} A maximum polynomial order of 4 with permutational
symmetry of 22222 is employed and this results in a total of 16,981 PIPs and thus linear coefficients. These PIP bases are generated using MSA software.\cite{NandiQuBowman2019,msachen}
The optimized coefficients are obtained by solving the above least-squares linear algebra (Eq. \ref{eq:5}) with the freely available FORTRAN code DGELSS.

\subsection{MULTIMODE Calculations}

Post-harmonic quantum methods based on vibrational self-consistent field (VSCF) and virtual-state configuration interaction (VCI) approaches have been known for almost fifty years. These methods have been implemented in our software called MULTIMODE. First, we present a brief recap of the  VSCF\cite{Bowman1978, vscf86} and VSCF/VCI scheme\cite{VCI} in MULTIMODE.\cite{CarterCulikBowman1997, Multimode2, BowmanCarterHuang2003} The computational code is based on the rigorous Watson Hamiltonian\cite{Watson1968} in mass-scaled normal coordinates, $\bm{Q}$, for non-linear molecules. This Hamiltonian is given by
\begin{equation}
\hat{H}=\frac{1}{2}\sum_{\alpha\beta}(\hat{J}_{\alpha}-\hat{\pi}_{\alpha})\mu_{\alpha\beta}(\hat{J}_{\beta}-\hat{\pi}_{\beta})-\frac{1}{2}\sum_k^F\frac{\partial^2}{\partial Q_k^2}-\frac{1}{8}\sum_{\alpha}\mu_{\alpha\alpha}+V(\bm{Q}),
\end{equation}
where $\alpha(\beta)$ represent the $x, y, z$ coordinates,
$\hat{J_{\alpha}}$ and $\hat{\pi_{\alpha}}$ are the components of the total and vibrational angular momenta respectively, $\mu_{\alpha\beta}$ is the
inverse of effective moment of inertia tensor, and $V(\bm{Q})$ is the full potential in terms of normal coordinates. The number of normal modes is denoted by $F$, and for non-linear molecules $F$ equals $3N-6$. In many applications of this Hamiltonian in the literature, the vibrational angular momentum terms are neglected and this approximation leads to an inaccurate result. Therefore, we include these terms in the MULTIMODE software.

In general, there are two major bottlenecks in applications to the VSCF/VCI scheme. One is the numerical evaluation of matrix elements (multi-dimensional integrals) and the second is the size of the H-matrix. Both naively have exponential dependence on the number of normal coordinates. An effective approach to deal with exponential scaling of matrix elements we represent the full potential in a hierarchical $n$-mode representation ($n$MR).\cite{CarterCulikBowman1997} In normal coordinates, this representation is given by
\begin{equation}
\begin{aligned}
& V(Q_1, Q_2, \cdots, Q_F)=\sum_iV_i^{(1)}(Q_i)+\sum_{i,j}V_{ij}^{(2)}(Q_i,Q_j)+\\
&\sum_{i,j,k}V_{ijk}^{(3)}(Q_i,Q_j,Q_k)
 +\sum_{i,j,k,l}V_{ijkl}^{(4)}(Q_i,Q_j,Q_k,Q_l)+\cdots,
\end{aligned}
\label{eq:nMR}
\end{equation}
where $V_i^{(1)}(Q_i)$ is the one-mode potential, i.e., the 1D cut through the full-dimensional PES in each mode, one-by-one,
$V_{ij}^{(2)}(Q_i,Q_j)$ is the intrinsic 2-mode potential among all pairs of modes, etc.  Here, intrinsic means that any $n$-mode term is zero if any of the arguments is zero. Also, each term in the representation is in principle of infinite order in the sense of a Taylor series expansion. So for example,  $V^{(1)}(Q)$ might look like a full Morse potential.

This representation has been used for nearly twenty years by a number of research groups; a sample of these are refs. \citenum{CarterCulikBowman1997, Multimode2, BowmanCarterHuang2003, RN12, RN13, ove12, konig15}.  It continues to be actively used in a variety of applications and theoretical developments.\cite{Rauhut21, Rauhut20a, Rauhut20b, nmodegpr, nmodetensor, ove2020} In MULTIMODE the maximum value of $n$ is 6.  However, from numerous tests it appears that a 4MR typically gives energies that are converged to within roughly 1--5 cm$^{-1}$\cite{bowman08, carter12, cart12}.  Thus we generally use 4MR with an existing full-dimensional PES and this is also done here.

The second major bottleneck to all VCI calculations is the diagonalization of the H-matrix, which as noted already can scale exponentially with the number of vibrational modes. This matrix results in the usual way following the VCI expansion of wavefunctions given in simplified notation by
\begin{equation}
\Psi_{L}=\sum\limits_{K} {c_K^{(L)}}\Phi_K,
\label{eqn:ci}
\end{equation}
where $\Phi_K$ are a complete, orthonormal set of functions. In the VSCF/VCI approach, these are the eigenfunctions of the VSCF Hamiltonian operator for the ground vibrational state. There are many strategies to deal with this. Basically, they all limit the size of the excitation space, with many schemes taken from electronic structure theory.
For example, the excitation space can be limited by using the hierarchical scheme of single, double, triple, etc. excitations. MULTIMODE uses this among other schemes and can consider up to quintuple excitations. A major difference with electronic structure theory is that the nuclear interactions go beyond 2-body.  This is immediately clear from the $n$-mode representation.  Thus, MULTIMODE tailors the excitation scheme for each term in this representation.  
Other schemes to prune the CI basis have been suggested and the reader is referred to reviews\cite{bowman08, vscf13, ove12, rauhut14, csa14, tenn16, Xiaohong2015,carr17,Rauhut21} for more details and specific details for the present calculations are given in the Electronic Supplementary Information (ESI). However, we note that in the present case, an iterative diagonalization routine is used to obtain the eigenvalues and eigenvectors of the H-matrix.  The eigenvalues are VSCF/VCI quantum vibrational energies, $E_L$ with corresponding eigenvectors, i.e., the expansion coefficients, $c_K^{(L)}$. 

In this paper, where the vibrational IR and power spectra play a central role, we make the following important remarks.  First, the quantum power spectrum is rigorously just the distribution of \textit{all} vibrational energies vs the vibrational energy.  This can simply be visualized as vertical sticks of say unit height at the energies $E_L$. This spectrum of energies is not the IR spectrum.  Indeed, this theoretical spectrum is virtually impossible to measure using IR and even IR and Raman spectroscopy.  The reason comes from well-known selection and propensity rules governing these spectroscopies.  In the present case, where low-resolution experimental IR spectra are compared with calculations the textbook selection ``$\Delta\nu$''=1 is assumed to hold. Of course, the IR spectrum can be calculated rigorously if the coordinate-dependent molecular dipole surface is available.  Unfortunately, that surface is not available.   
The second remark is how we calculate a quantum power spectrum that can be reasonably compared to the IR spectrum and to the other calculated power spectra (more comments about these are given below). The approach we take is to examine the expansion coefficients for all the quantum states obtained with the vibrational bands of interest here, namely the OH- and the CH-stretch bands.  We filter out just those states with dominant expansion coefficient(s) for one quantum of excitation in the OH-stretch or the CH-stretch. Thus, we replace the unit stick at each energy with the square of the VCI coefficient corresponding to a CH-stretch for that band and the square of the VCI coefficient corresponding to a OH-stretch for that band. Therefore, we can approximate the Intensity, $I(E_{CH / OH}) \propto |C_{(CH / OH)}|^2$. 
  Finally, these sticks are Gaussian broadened and then Boltzmann weighted according to the relative population of a given conformer at 300 K.  (Note, broadening of sticks is typically done for pure vibrational spectra, where the total angular momentum is zero and thus thermal broadening from rotational populations and selections rules are absent.) So, the working formula to estimate intensity is

\begin{equation}
I(E^{Conf\{i\}}_{CH / OH}) \propto wt^{Conf\{i\}} \cdot |C^{Conf\{i\}}_{(CH / OH)}|^2; \forall \{i\} \rightarrow 1, 2,. . ., 5 .
\end{equation}
\begin{equation}
I(E^{Conf\{i\}}_{CH / OH}) \approx I(constant) \cdot wt^{Conf\{i\}} \cdot |C^{Conf\{i\}}_{(CH / OH)}|^2
\label{eqn:Int_QM}
\end{equation}

\noindent where $I(constant)$ is the arbitrary constant value, $wt^{Conf\{i\}}$ is the corresponding Boltzmann weight. This Eq. \ref{eqn:Int_QM} is followed to estimate the intensity of CH- or OH-stretches for each conformer from MULTIMODE calculations.

\subsection{AS-SCIVR Calculations}

The adiabatically switched semiclassical initial value representation (AS SCIVR) is a recently developed two-step semiclassical approach\cite{Conte_Ceotto_ASSCIVR_2019,Botti_Conte_ASotf_2022} able to regain quantum effects starting from classical trajectories. It differs from standard semiclassical techniques\cite{miller2001semiclassical,Huber_Heller_SCdynamics_1987} in the way the starting conditions of the semiclassical dynamics run are selected. In AS SCIVR a preliminary adiabatic switching\cite{as2016} dynamics is performed, a procedure not present in previous semiclassical techniques. This allows one to start from an approximate true quantization of the initial conditions. Therefore, the exit atomic positions and momenta of the adiabatic switching run serve as starting conditions for the subsequent semiclassical dynamics trajectory.

The adiabatic switching Hamiltonian is\cite{Qiyan_Gazdy_AdiabaticSwitching_1988,Saini_Taylor_AdiabaticSwitching_1988,Nagy_Lendvay_AdiabaticSwitching_2017}

\begin{equation}
    H_{\mathrm{as}} = \left[ 1-\lambda (t) \right] H_{\mathrm{harm}} + \lambda (t) H_{\mathrm{anh}},
    \label{eqn:switching_ham}
\end{equation}
where $\lambda (t)$ is the following switching function
\begin{equation}
    \lambda(t) = \frac{t}{T_{\mathrm{AS}}} - \frac{1}{2\pi} \sin \left( \frac{2\pi t}{T_{\mathrm{AS}}} \right),
    \label{eqn:switching}
\end{equation}
$H_{harm}$ is the harmonic Hamiltonian built from the harmonic frequencies of vibration, and $H_{anh}$ is the actual molecular vibrational Hamiltonian. In our simulations $T_{AS}$ has been chosen equal to 25000 a.u. (about 0.6 ps) and time steps of 10 a.u. have been employed. 4000 trajectories were evolved in each AS-SCIVR calculation. The AS-SCIVR zero-point energy estimate has been obtained by starting the AS run with no quanta of excitation in the modes, i.e. from the harmonic zero-point energy. Conversely, AS-SCIVR estimates of C-H and O-H stretches have been obtained by starting the AS run with an additional quantum of harmonic excitation to the specific mode under investigation. AS-SCIVR calculations have been performed for the global minimum geometry.

Once the adiabatic switching run is over, the trajectories are evolved according to $H_{anh}$ for another 25000 a.u. with the same step size to collect the dynamical data needed for the semiclassical calculation. This relies on Kaledin and Miller's time-averaged version of semiclassical spectroscopy. Therefore, the working formula is 
\begin{equation}
\begin{split}
 I_{as} (E) = & \left( \frac{1}{2\pi \hbar} \right)^{N_{v}} \sum_{i=1}^{N_{traj}}\frac{1}{2\pi\hbar T}  \\ & \left\vert \int_{0}^{T} \, dt e^{\frac{i}{\hbar}\left[ S_{t} (\mathbf{p}_{as},\mathbf{q}_{as}) + Et + \phi_{t} (\mathbf{p}_{as},\mathbf{q}_{as})\right ]} 
\langle \Psi (\mathbf{p}_{eq},\mathbf{q}_{eq}) \vert g (\mathbf{p}^{\prime}_{t},\mathbf{q}^{\prime}_{t}) \rangle \right\vert^{2},
    \label{eqn:astascivr}
\end{split}    
\end{equation}
where $I_{as} (E )$ indicates that a vibrational spectral density is calculated as a function of the vibrational energy $E$. 
In Eq. \ref{eqn:astascivr}, $N_v$ is the number of vibrational degrees of freedom of the system, i.e. 24 in the case of ethylene glycol. $T$ is the total evolution time of the dynamics for the semiclassical part of the simulation. As anticipated, we chose $T$ equal to 25000 a.u. with a time step size of 10 a.u. $({\bf p}_{t}^\prime,{\bf q}_{t}^\prime)$ is the instantaneous full-dimensional phase-space trajectory started at time 0 from the final phase space condition $({\bf p}_{as},{\bf q}_{as})$ of the adiabatic switching part of the simulation. $S_t$ is the classical action along the semiclassical trajectory, and $\phi_t$ is the phase of the Herman-Kluk pre-exponential factor based on the elements of the stability matrix and defined as 
\begin{equation}
    \phi_{t} = \mathrm{phase} \left[ \sqrt{\left\vert \frac{1}{2} \left( \frac{\partial \mathbf{q}^\prime_{t}}{\partial \mathbf{q}_{as}} + \Gamma^{-1} \frac{\partial \mathbf{p}^\prime_{t}}{\partial \mathbf{p}_{as}} \Gamma -i\hbar \frac{\partial \mathbf{q}^\prime_{t}}{\partial \mathbf{p}_{as}} \Gamma + \frac{i \Gamma^{-1}}{\hbar} \frac{\partial \mathbf{p}^\prime_{t}}{\partial \mathbf{q}_{as}} \right) \right\vert} \right],
    \label{eqn:phase}
\end{equation}
where $\Gamma$ is an $N_v\times N_v$ matrix usually chosen to be diagonal with elements numerically equal to the harmonic frequencies. 

Classical chaotic dynamics can lead to numerical inaccuracies in the semiclassical propagation, so, following a common procedure in semiclassical calculations, we have rejected the trajectories based on a 1\% tolerance threshold on the monodromy matrix determinant value. In the case of ethylene glycol this led to a rejection rate between 75 and 80\% of trajectories.  Finally, the working formula is completed by a quantum mechanical overlap between a quantum reference state $|\Psi\rangle$ and a coherent state $|g\rangle$ characterized by the following representation in configuration space 
\begin{equation}
    \langle\mathbf{q}|g(\mathbf{p}^{\prime}_{t},\mathbf{q}^{\prime}_{t})\rangle = \left( \frac{\det (\Gamma)}{\pi^{N_{\nu}}} \right) \exp \left\lbrace - (\mathbf{q}-\mathbf{q}^\prime_{t})^{T} \frac{\Gamma}{2} (\mathbf{q}-\mathbf{q}^\prime_{t}) + \frac{i}{\hbar}\mathbf{p}^{\prime T}_{t}(\mathbf{q}-\mathbf{q}^\prime_{t}) \right\rbrace.
    \label{eqn:coherent_superposition}
\end{equation}
The reference state $|\Psi\rangle$ is usually chosen to be itself a coherent state. In Eq. (\ref{eqn:astascivr}) $|\Psi\rangle$ is written as $|\Psi (\mathbf{p}_{eq},\mathbf{q}_{eq})\rangle$, where ${\bf p}_{eq}$ stands for the linear momenta obtained in harmonic approximation setting the geometry at the equilibrium one (${\bf q}_{eq}$).

Finally, we stress that this approach to the power spectrum is general and powerful, but it does not in practice resolve the quantum power spectrum discussed above in the way MULTIMODE does. The input for each band is a semiclassical approximation to a fundamental excitation in the OH and CH-stretching modes. So the dominant signal is for these excitations. The method can in principle also resolve weak combination bands at higher energies built on the fundamental transition as well as Fermi resonances. This has been done in the present case and illustrated in the Results and Discussion section. For the sake of completeness, we mention that more elaborated semiclassical approaches have been developed to decompose each anharmonic semiclassical signal (i.e. wavefunction) into its harmonic components.\cite{Micciarelli_Ceotto_SCwavefunctions_2018,Aieta_Ceotto_nuclear_densities_2020} This procedure would require more work but would also be more directly comparable to MULTIMODE calculations providing better resolved quantum power spectra. However, while MULTIMODE gets all states at once upon diagonalization of an H matrix, SC methods need to focus to one state at a time.  

Finally, we note that since AS-SCIVR is a semi-classical method that aims for quantization of the vibrational modes, and notably those corresponding to the OH-stretch and CH-stretch, the expectation is that it will perform well here.  Such quantization is absent in MD calculations as well as the the TRPMD ones.  The MD calculations at best capture some anharmonicity of these modes; however, at 300 K, this is not really achieved given the ``stiffness" of these modes.  TRPMD aims to capture anharmonicity visited by zero-point motion, and this is expected and seen to provide a more realistic description of these modes.  However, quantization is also absent in TRPMD and so the anharmonicity associated with the vibrationally excited states is not captured by TRPMD.  The extent of the missing anharmonicity in both classical and TRPMD calculations and the accuracy of AS-SCIVR is of course problem-specific and an objective here is to quantify this for ethylene glycol.

\section{Results and discussion}

\subsection{PES Fitting}
\begin{figure}
    \centering
    \includegraphics[width=0.8\linewidth]{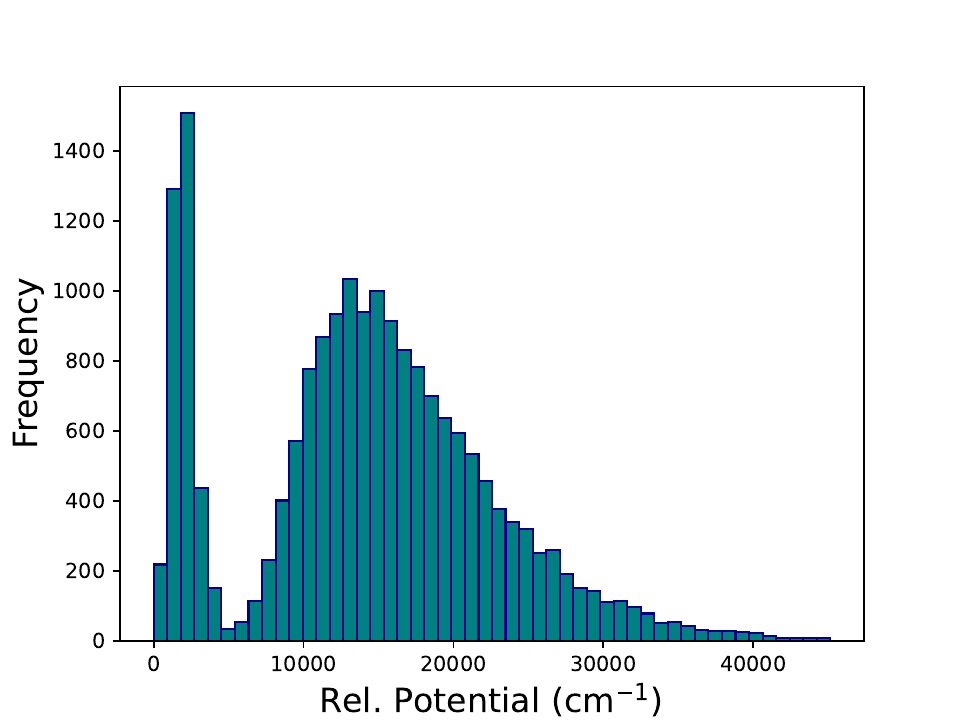}
    \caption{MP2/aug-cc-pVTZ energy distribution of the data set with respect to the minimum energy.}
    \label{fig:hist}
\end{figure}

A full-dimensional PES of ethylene glycol has been developed using the PIP approach and this PES has been employed for all results presented in this section. A total of 18,772 geometries have been employed in developing this PES and the dataset has been taken from a recently reported article.\cite{Sai_2024} 
All electronic energy calculations were performed using the MP2/aug-cc-pVTZ  (MP2/aVTZ) level of theory, as described previously.\cite{Sai_2024} The distribution of these energies is shown in Figure \ref{fig:hist}.  As seen, there is a concentration of energies between 0 and roughly 5,000 \cm.  These are used to establish a ``reaction surface'', where the potential is minimized with respect to three dihedral angles, as described in ref. \citenum{Sai_2024}.  Energies for displaced configuration from this minimum surface constitute the second broad distribution of energies.  Potential gradients were also reported for these energies and these constitute an additional 563,160 pieces of data.

\begin{figure}
    \centering
    \includegraphics[width=0.8\linewidth]{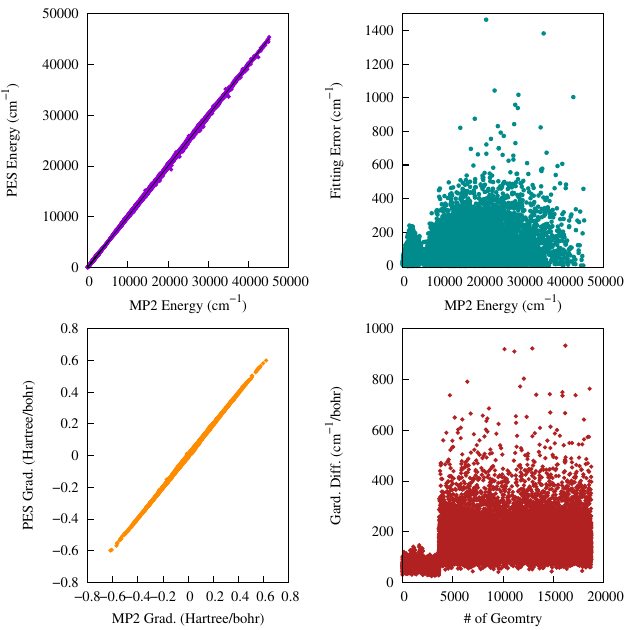}
    \caption{Left upper panel shows the direct MP2 vs PES energies for the training data set relative to the MP2 minimum energy. Corresponding fitting errors relative to the minimum energy are shown in the right upper panels.
    The left lower panel represents the correlation between the direct MP2 and PES gradients for the training dataset. Corresponding gradient fitting errors are shown in the lower right panels.}
    \label{fig:corre}
\end{figure}

The PIP basis to fit this PES is generated using MSA software.\cite{NandiQuBowman2019,msachen}
 We perform both weighted average and unweighted fitting for this PES. In the process of weighted average fitting, a weight is assigned to each data point based on its energy. The weight is given by $wt=E_0/(E_0+dE)$, where $dE$ is the energy relative to the minimum in a.u., and $E_0$ is the parameter that we could modify. For the unweighted fitting, $E_0$ is typically set as a large number, such as $10^{10}$~a.u., resulting in all weights essentially being 1. 
Here we have used $E_0$ as 0.01 to get the weighted average fitting.
The RMSEs for the unweighted and weighted fitting are 124~cm$^{-1}$ and 70~cm$^{-1}$ for energies; 0.0009617 Hartree/bohr and 0.0005455 Hartree/bohr for forces, respectively. We used the weighted fitted PES for all the studies. A correlation plot between the weighted PES energies and gradients vs corresponding direct MP2 energies and gradients along with the absolute fitting errors for the 18,722 data points is shown in Figure. \ref{fig:corre}.

\begin{figure}
    \centering
    \includegraphics[width=0.8\linewidth]{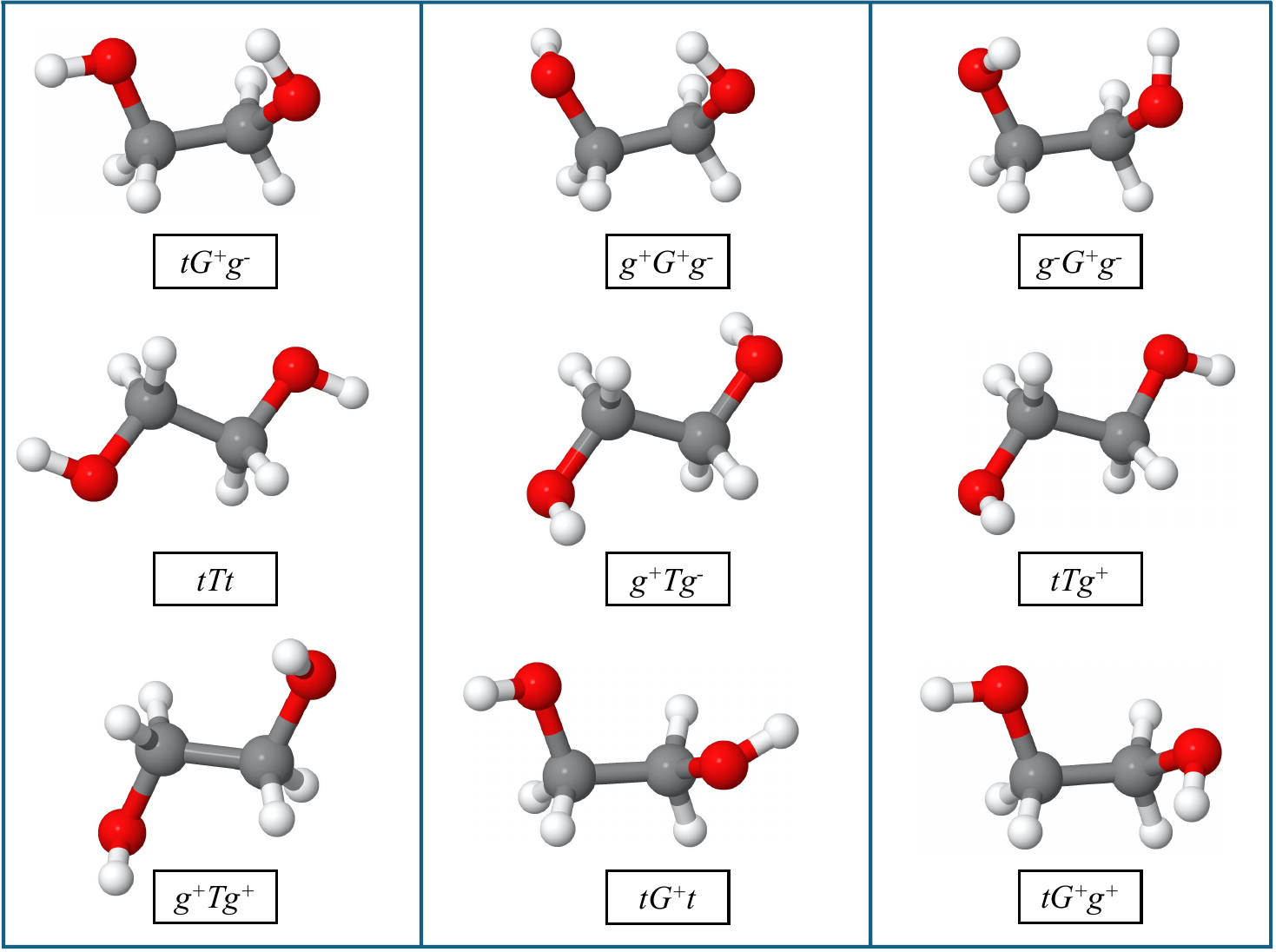}
    \caption{PES optimized geometry of nine low-lying conformers of ethylene glycol.}
    \label{fig:geom}
\end{figure}

To examine the standard fidelity of this PES, we first perform geometry optimizations of ten low-lying conformers of ethylene glycol. The structures of nine low-lying conformers of ethylene glycol are shown in Figure \ref{fig:geom}.
A comparison of the relative energetics of these conformers is shown in Table. \ref{tab:Energetics}. It is seen that PES-optimized conformers perfectly preserved the energy order in accord with the direct MP2/aVTZ energies as well as CCSD(T) ones and also PES-optimized energies are within 30 - 60 cm$^{-1}$ of the direct MP2/aVTZ energies. 
Next, to examine the vibrational frequency predictions of the PES, we perform normal-mode analyses for five low-lying conformers. The comparison of harmonic mode frequencies for these five conformers with direct MP2/aVTZ ones is shown in Table. \ref{tab:Freq_wt}.
The agreement with the direct MP2/aVTZ frequencies for these conformers is overall very good; the maximum error is 45 cm$^{-1}$ for the lowest frequency
mode of \textbf{\ce{g+Tg-}} conformer, but most of the frequencies are
within a few cm$^{-1}$ of the \textit{ab initio} ones, especially the high-frequency \ce{O-H} and \ce{C-H} stretches are within 20 - 25 cm$^{-1}$ of direct MP2 values. And the mean absolute errors (MAEs) are within 12 cm$^{-1}$.
All these local minima are confirmed by obtaining no imaginary frequency except for the \textbf{\ce{g- G+g-}} conformer, where we obtained a small imaginary frequency of 67$i$. This imaginary mode corresponds to the large amplitude torsional mode where the potential is very shallow. 

\begin{table}[h]
\scriptsize
    \centering
    \begin{tabular}{ccccccccccc} 
    \toprule
    \toprule
    \textbf{Method} & \textbf{\ce{tG+g-}} & \textbf{\ce{g+G+g-}} & \textbf{\ce{g- G+g-}} & \textbf{tTt} & \textbf{\ce{g+Tg-}} & \textbf{\ce{tTg+}} & \textbf{\ce{g+Tg+}} & \textbf{\ce{tG+t}} & \textbf{\ce{tG+g+}} & \textbf{cCt} \\
    \toprule
    \textbf{MP2/aVTZ$^a$} & 0.0 & 149 & 330 & 912 & 979 & 971 & 1046 & 1084 & 1249 & 2298 \\
    \textbf{CCSD(T)$^a$} & 0.0 & 113 & 305 & 876 & 908 & 919 & 968 & 1093 & 1210 & 2324 \\
    \textbf{PES} & 0.0 & 69 & 290 & 905 & 1015 & 904 & 972 & 1115 & 1219 & 2252 \\
    \textbf{PES$^{wt}$} & 0.0 & 79 & 292 & 912 & 1018 & 894 & 990 & 1111 & 1215 & 2261 \\
    \toprule
    \end{tabular}
    \\
    $^a$ From Table S-1 in Ref. \citenum{Sai_2024}  \\
    \caption{Relative energetics (cm$^{-1}$) of ten low-lying conformers of ethylene glycol with respect to the global minimum.}
    \label{tab:Energetics}
\end{table}

\begin{table}[h]
\scriptsize
    \centering
    \begin{tabular}{ccccccccccc} 
    \toprule
    \toprule
    \textbf{Mode} &  \multicolumn{2}{c}{\textbf{\ce{tG+g-}}} & \multicolumn{2}{c}{\textbf{\ce{g+G+g-}}} & \multicolumn{2}{c}{\textbf{\ce{g- G+g-}}} & \multicolumn{2}{c}{\textbf{tTt}} & \multicolumn{2}{c}{\textbf{\ce{g+Tg-}}} \\
    \cmidrule(rl){2-3} \cmidrule(rl){4-5} \cmidrule(rl){6-7} \cmidrule(rl){8-9} \cmidrule(rl){10-11}
    & MP2/TZ$^a$ & PES & MP2/TZ$^a$ & PES & MP2/TZ$^a$ & PES & MP2/TZ$^a$ & PES & MP2/TZ$^a$ & PES \\
    \toprule
    1 & 168 & 170 & 168 & 167 & 100 & 67i & 116 & 132 & 141 & 145 \\
    2 & 247 & 216 & 292 & 301 & 159 & 167 & 217 & 190 & 250 & 192 \\
    3 & 329 & 332 & 327 & 326 & 321 & 326 & 230 & 209 & 268 & 223 \\
    4 & 420 & 407 & 452 & 459 & 428 & 432 & 291 & 295 & 296 & 285 \\
    5 & 523 & 529 & 536 & 531 & 528 & 530 & 481 & 483 & 475 & 468 \\
    6 & 887 & 889 & 878 & 874 & 881 & 880 & 839 & 842 & 803 & 787 \\
    7 & 904 & 905 & 897 & 900 & 885 & 889 & 1009 & 1046 & 1027 & 1043 \\
    8 & 1066 & 1071 & 1059 & 1061 & 1051 & 1058 & 1076 & 1074 & 1074 & 1074 \\
    9 & 1100 & 1098 & 1073 & 1073 & 1063 & 1073 & 1094 & 1090 & 1090 & 1083 \\
    10 & 1130 & 1135 & 1122 & 1121 & 1126 & 1123 & 1167 & 1168 & 1109 & 1107 \\
    11 & 1178 & 1207 & 1204 & 1215 & 1198 & 1200 & 1190 & 1236 & 1140 & 1154 \\
    12 & 1269 & 1277 & 1246 & 1250 & 1257 & 1280 & 1235 & 1241 & 1320 & 1337 \\
    13 & 1296 & 1314 & 1374 & 1360 & 1383 & 1401 & 1288 & 1306 & 1339 & 1343 \\
    14 & 1384 & 1390 & 1377 & 1379 & 1389 & 1403 & 1319 & 1328 & 1370 & 1364 \\
    15 & 1420 & 1429 & 1406 & 1405 & 1404 & 1414 & 1409 & 1427 & 1404 & 1381 \\
    16 & 1455 & 1465 & 1435 & 1437 & 1422 & 1428 & 1487 & 1513 & 1433 & 1429 \\
    17 & 1516 & 1519 & 1511 & 1514 & 1508 & 1513 & 1541 & 1545 & 1521 & 1507 \\
    18 & 1524 & 1525 & 1521 & 1519 & 1512 & 1526 & 1551 & 1549 & 1535 & 1535 \\
    19 & 3053 & 3056 & 3026 & 3027 & 3071 & 3069 & 3057 & 3065 & 3067 & 3067 \\
    20 & 3058 & 3064 & 3070 & 3073 & 3074 & 3077 & 3064 & 3066 & 3076 & 3075 \\
    21 & 3114 & 3118 & 3134 & 3136 & 3140 & 3142 & 3102 & 3104 & 3124 & 3127 \\
    22 & 3149 & 3148 & 3159 & 3160 & 3148 & 3150 & 3127 & 3130 & 3150 & 3149 \\
    23 & 3808 & 3831 & 3794 & 3815 & 3845 & 3856 & 3857 & 3870 & 3840 & 3856 \\
    24 & 3856 & 3871 & 3831 & 3847 & 3846 & 3876 & 3858 & 3878 & 3842 & 3869 \\
    \toprule
    \bf{MEA} & & \bf{8} & & \bf{5} & & \bf{8} & & \bf{12} & & \bf{12} \\
    \toprule
    \end{tabular}
    \\
    $^a$ From Table S-3 in Ref. \citenum{Sai_2024}\\
    \caption{Normal mode frequencies (cm$^{-1}$) for five low-lying conformers of ethylene glycol from weighted fitting PES.}
    \label{tab:Freq_wt}
\end{table}

Normal modes 19-22 correspond to the four CH-stretches and  23 and 24 correspond to the two OH-stretches.  Of the various bending modes, the highest frequency ones, modes 17 and 18, are of interest as they are roughly in the ratio 1:2 with the CH-stretch modes at 3053 and 3058 \cm.  We return to this in the next section.

\subsection{MULTIMODE Results}

As a quantum nuclear application of the PES, we performed VSCF/VCI calculations using Version 5.1.4 of MULTIMODE.\cite{Bowman1978, CarterCulikBowman1997,BowmanCarterHuang2003} For all the calculations, a four-mode representation of the potential in mass-scaled normal coordinates and a two-mode representation of the effective inverse moment of inertia for the vibrational angular momentum terms in the exact Watson Hamiltonian are used.\cite{Watson1968}. The formalism is based on the configuration interaction (CI) approach from the virtual space of the ground vibrational state VSCF Hamiltonian. Here we explore reduced-mode coupling models, i.e., 15-mode models, where these sets of modes start with the highest frequency OH-stretches and proceed in decreasing frequency.  In this case, the maximum mode combination excitations are 10 10 10 8, which means that singles through triple excitations extend to a maximum sum of quanta of 10, and for quadruple excitations, the maximum is 8. This excitation space leads to the VCI H-matrix of order 155 026 for the 15-mode calculation. We compute 200 CI vibrational states up to the energy of 4000 cm$^{-1}$.

MULTIMODE calculations were performed for the five low-lying conformers. Table \ref{tab:CI_Freq1} shows MULTIMODE VSCF/VCI frequencies with the corresponding harmonic ones and the three largest VCI coefficients in the expansion basis above for the global minimum conformer (\textbf{\ce{tG+g-}}). Results for other conformers are given in the ESI.  First, note that the harmonic frequencies are noticeably overestimated compared to the corresponding CI values, particularly for the  \ce{CH}- and \ce{OH}-stretches, which are overestimated by approximately 200 cm$^{-1}$. This highlights the impact of anharmonicity, as expected. The presence of mixing states, notably due to Fermi resonances, is also observed. This similar trend we also observed for the other four conformers, and the VCI frequencies with the three largest VCI coefficients are provided in Tables S2-5 in ESI. 

\begin{table}
    \centering
    \begin{tabular}{ccccc} 
    \toprule
    \toprule
    \textbf{Mode} & \textbf{Har. Freq.} & \textbf{CI Freq.} & \textbf{VCI Coeff.} & \textbf{Corresponding Modes} \\
    \toprule
    10 & 1135 & 1119 & 0.9957 & $\nu_{10}$ \\
    11 & 1207 & 1167 & -0.9889, 0.0581, 0.0455  & $\nu_{11}$, $\nu_{12}$, $\nu_{14}$ \\
    12 & 1277 & 1246 &  0.9774, 0.1745, 0.0632  & $\nu_{12}$, $\nu_{13}$, $\nu_{11}$ \\
    13 & 1314 & 1268 & -0.9633, 0.1771, 0.1318  & $\nu_{13}$, $\nu_{12}$, $\nu_{15}$ \\
    14 & 1390 & 1340 &  0.9344, 0.3043, 0.1418  & $\nu_{14}$, $\nu_{15}$, $\nu_{13}$ \\
    15 & 1429 & 1387 & -0.9354, 0.3206, -0.0952 & $\nu_{15}$, $\nu_{14}$, $\nu_{13}$ \\
    16 & 1465 & 1429 & -0.9913, -0.0475, 0.0418 & $\nu_{16}$, $\nu_{18}$, $\nu_{15}$ \\
    17 & 1519 & 1472 &  0.9552, -0.2789, -0.0396 & $\nu_{17}$, $\nu_{18}$, $\nu_{15}$ \\
    18 & 1525 & 1478 &  0.9549, 0.2799, -0.0395 & $\nu_{18}$, $\nu_{17}$, $\nu_{16}$ \\
    \\
    \multirow{3}{*}{19} & \multirow{3}{*}{3056} & 2798 &  0.8310, -0.3908, 0.2718 & ($\nu_{15}$ + $\nu_{16}$), $\nu_{19}$, 2$\nu_{16}$ \\
    & & 2918 & 0.5991,  0.4551, 0.3765 & $\nu_{19}$, 2$\nu_{18}$, 2$\nu_{16}$ \\
    & & 2972 & 0.6764, -0.3948, 0.3448 & $\nu_{21}$, 2$\nu_{18}$, $\nu_{19}$ \\
    & & 2981 & -0.5441, 0.5289, -0.4209 & ($\nu_{17}$ + $\nu_{18}$), $\nu_{21}$, $\nu_{19}$ \\
    \\
    \multirow{2}{*}{20} & \multirow{2}{*}{3064} & 2836 & -0.8092, 0.3438, -0.2593 & 2$\nu_{16}$, ($\nu_{15}$ + $\nu_{15}$), $\nu_{20}$ \\
    & & 2906 & -0.8184, 0.2904, -0.2245 & ($\nu_{16}$ + $\nu_{17}$), $\nu_{20}$, ($\nu_{17}$ + $\nu_{18}$) \\
     &  & 2915 & 0.5181, 0.4893, 0.4766 & ($\nu_{16}$ + $\nu_{17}$), 2$\nu_{17}$, $\nu_{20}$ \\
     & & 2961 & 0.5192, 0.4968, 0.4057 & $\nu_{20}$, ($\nu_{17}$ + $\nu_{18}$), 2$\nu_{18}$ \\
     \\
     \multirow{2}{*}{21} & \multirow{2}{*}{3118} & 2972 & 0.6764, -0.3948, 0.3448 & $\nu_{21}$, 2$\nu_{18}$, $\nu_{19}$ \\
      & & 2981 & -0.5441, 0.5289, -0.4209 & ($\nu_{17}$ + $\nu_{18}$), $\nu_{21}$, $\nu_{19}$ \\
      \\
      22 & 3148 & 3012 & -0.8881, -0.1973,  0.1762 & $\nu_{22}, 2\nu_{17}, (\nu_{17} + \nu_{18})$ \\
      23 & 3831 & 3629 & -0.9645, -0.0877, -0.0796 & $\nu_{23}, (\nu_{11} + 2\nu_{13}), (2\nu_{10} + \nu_{15})$ \\
      24 & 3871 & 3681 &  0.9600, -0.1649,  0.0869 & $\nu_{24}, (\nu_{11} + 2\nu_{13}), \nu_{23}$ \\
    \toprule
    \end{tabular}
    \caption{VSCF/VCI Energies (cm$^{-1}$) and VCI expansion coefficients for \textbf{\ce{tG+g-}} conformer.}
    \label{tab:CI_Freq1}
\end{table}

The VCI coefficients for states with energies above 1478 \cm ~ are the ones of interest for the calculation of the power spectrum, according to the remarks above. As seen, the states in the region of the CH-stretch are strongly mixed with overtones and combination bands of the lower-frequency bends. The state at 2798 \cm ~ is dominantly the combination band $\nu_{15}$+$\nu_{16}$ with a VCI weight of 0.16 for the CH-stretch. By contrast the states for the OH-stretches, modes 23 and 24 are ``pure'', i.e., with VCI coefficients of 0.96 in magnitude.

Next, we present the central results of this paper, namely comparisons between theory and experiment for the vibrational spectra of the OH and CH-stretch bands.  

\subsubsection{OH-Stretch}
First, we present the spectra for the OH-stretching modes. We obtained anharmonic OH stretching frequencies from MULTIMODE calculations as 3629 and 3681 cm$^{-1}$ for the global minimum conformer (\textbf{\ce{tG+g-}}), whereas the harmonic ones are 3831 and 3871 cm$^{-1}$ (From Table \ref{tab:CI_Freq1}). Anharmonic OH stretching frequencies of the other four low-lying conformers are provided in Tables S2-5 in ESI.

\begin{figure}[H]
    \centering
    \includegraphics[width=1.0\linewidth]{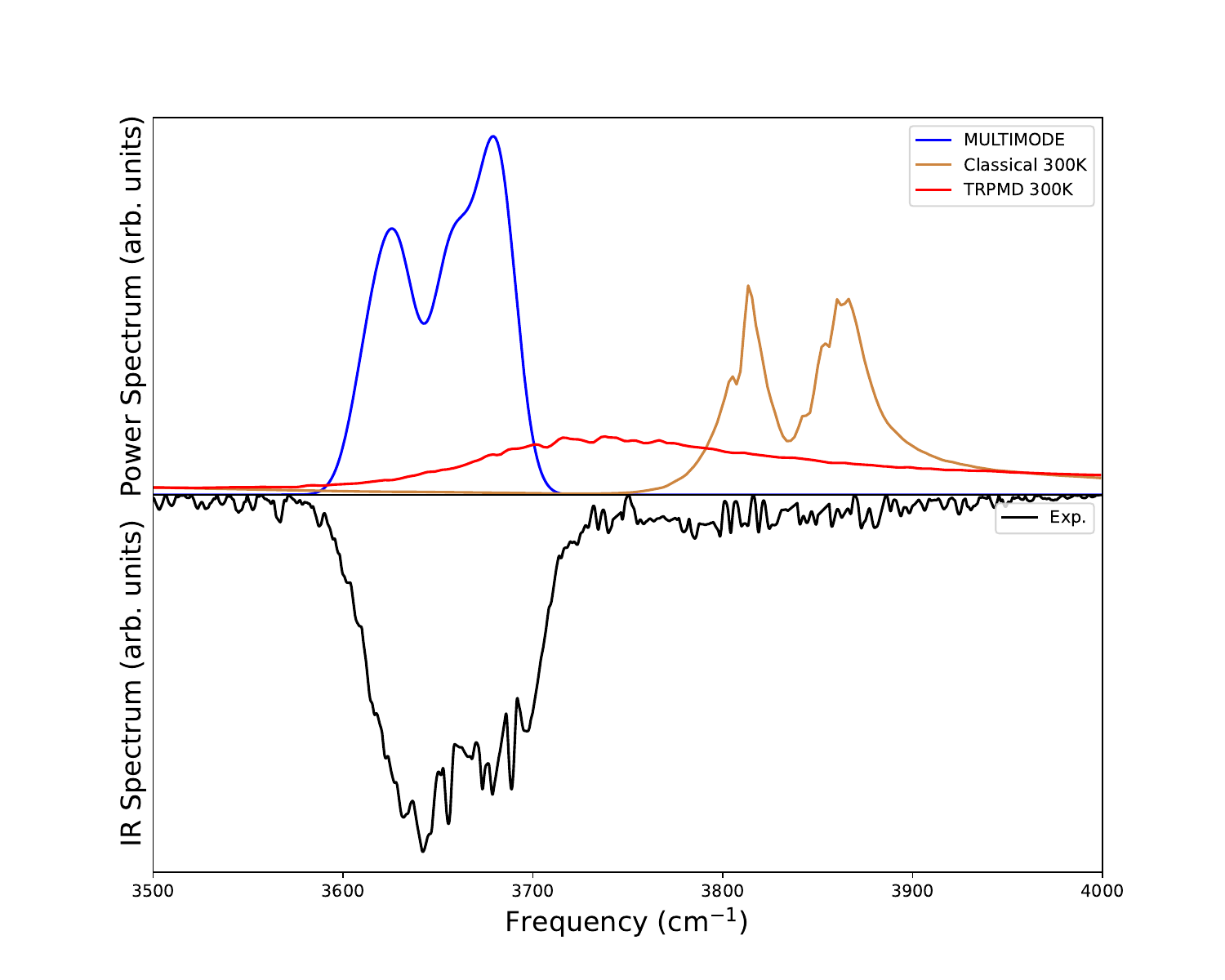}
    \caption{MULTIMODE spectra (blue curve) computed from VSCF/VCI calculations (with Boltzmann-weighted and smoothing by Gaussian broadening), power spectra at 300 K from Classical MD (orange curve) and TRPMD (red curve) simulations from Arandhara \textit{et al}\cite{Sai_2024}compared with experimental IR spectra (black curve) from Das \textit{et al}\cite{das2015} for \ce{OH} stretching region. See text for details.}
    \label{fig:freq_OH}
\end{figure}

The leading expansion coefficients are equal to 0.9 or greater for the OH-stretch VSCF/VCI basis function, as shown in Table \ref{tab:CI_Freq1} and as noted already. Thus, from simple zero-order arguments we expect the power and IR spectra to be quite similar for this fundamental transition. And indeed that is seen. 
A comparison of anharmonic \ce{OH} stretching frequencies with the experimental one and the TRPMD and classical MD power spectra are shown in Figure \ref{fig:freq_OH}. As we don't have a dipole moment surface of ethylene glycol, obtaining the exact intensities of these corresponding eigenstates is impossible. Therefore, we first try to make a stick plot by taking all anharmonic eigenstates of \ce{OH} stretching (obtained from MULTIMODE calculations) for five low-lying conformers and assign an arbitrary intensity of 0.2 for each eigenstate. Then we make thermal averaging of these sticks by multiplying each eigenstate by its corresponding Boltzmann weight. The dotted sticks in Figure S1 in the ESI represent the thermally average stick plot of the anharmonic OH stretching and making it more realistic spectra we apply Gaussian broadening denoted by a blue line. 

As seen, the VSCF/VCI power spectrum aligns excellently with the experimental IR spectrum. The classical MD spectrum is in poor agreement with the experiment for this strongly anharmonic band.  This is expected, since the peaks in the MD spectrum basically aligns with the harmonic OH-stretch energies.  Finally, while the very broad TRPMD band does exhibit some down-shift anharmonicity, it still overestimates the experimental band peak by about 100 cm$^{-1}$

\subsubsection{CH-Stretch}
Next, we consider the CH-stretch band. As can be surmised from the detailed results shown in Table \ref{tab:CI_Freq1}, this band is not as ``simple'' as the OH-stretch one, owing to large Fermi mixing among the basis states.  And, as a result, larger differences between the power and IR spectra are expected, owing to the likely strong variation in IR intensity across the band.  We defer a discussion of these resonances and their absence in the TRPMD and classical MD  simulations to the Discussion section. 

\begin{figure}[H]
    \centering
    \includegraphics[width=0.9\linewidth]{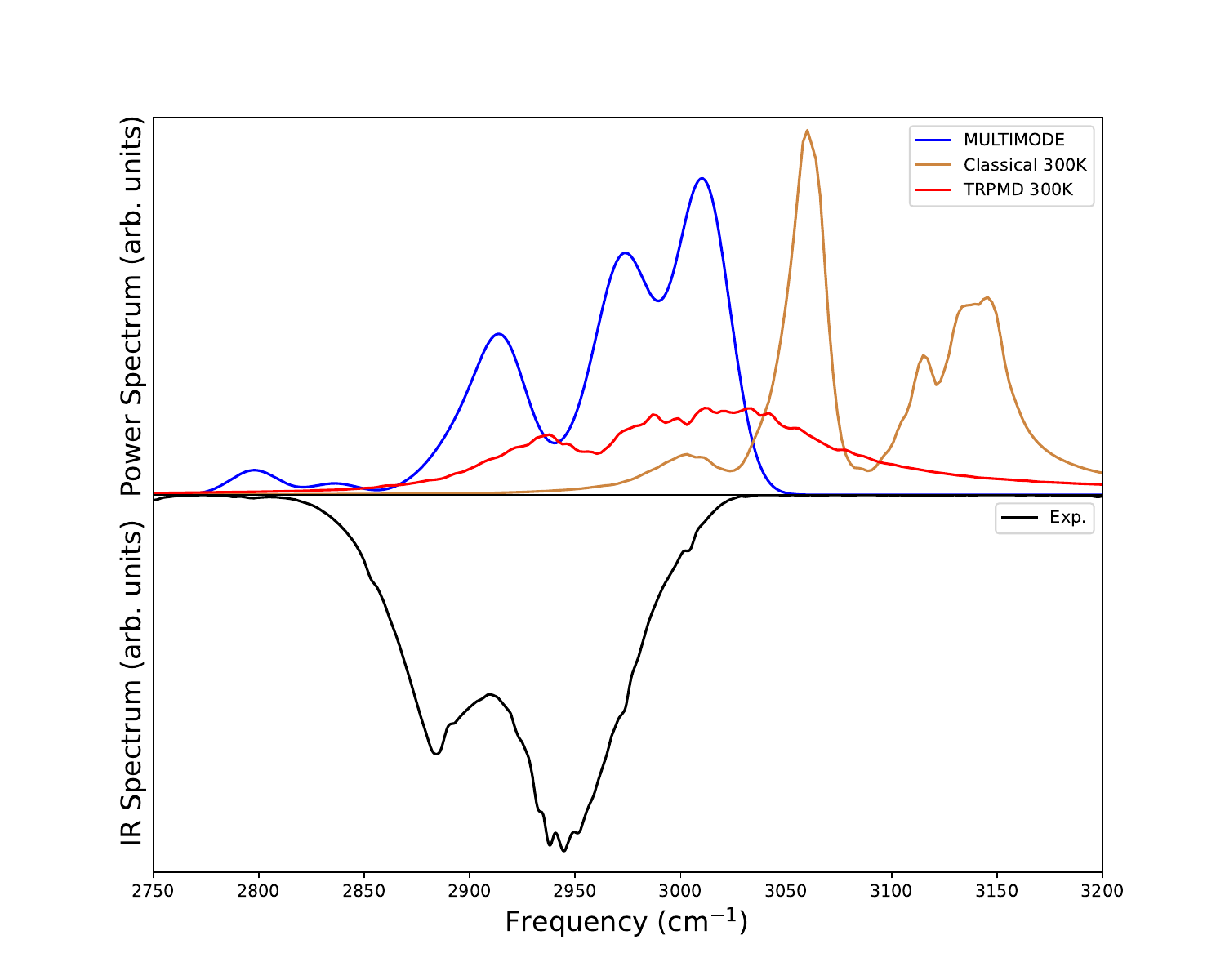}
    \caption{ MULTIMODE spectra (blue curve) computed from VSCF/VCI calculations, power spectra at 300 K from Classical MD (orange curve) and TRPMD (red curve) simulations from Arandhara \textit{et al}\cite{Sai_2024}compared with experimental IR spectra (black curve) from Das \textit{et al}\cite{das2015} for \ce{C-H} stretching region. See text for details, especially for the absence of the MULTIMODE band at 2800 \cm.}
    \label{fig:freq_CH}
\end{figure}

With the above remarks in mind consider the spectral results shown in Figure \ref{fig:freq_CH} (the corresponding stick plot of this C-H band is shown in Figure S2 in the ESI). As seen, the present VSCF/VCI band is closer to the experimental one than the TRPMD and MD bands, which are upshifted from the experiment by roughly 40 and 80 \cm, respectively. The VSCF/VCI band at roughly 2800 \cm~ is evidently absent in the experimental IR spectrum. This can be explained by examining the results of Table \ref{tab:CI_Freq1} for this band. As seen this lowest energy ``CH-stretch'' is a strongly mixed state, with the leading VCI coefficient corresponding to the combination band $\nu_{15}$+$\nu_{16}$ of two bends. Indeed, the sum of the VSCF/VCI energies of these bends equals 2816 \cm which is quite close to the eigenstate energy of 2798 \cm. ~So this band is a combination band, which from elementary considerations is expected to have much smaller IR intensity than a fundamental CH-stretch.  Thus, its absence in the experimental IR spectrum is not surprising.

As seen for the OH-stretch, the VSCF/VCI power spectrum aligns much better with the experimental IR spectrum than the classical and TRPMD spectra. The CH-stretch is complex, as noted, owing to multiple resonance interactions with bending modes.

Before presenting the AS-SCIVR results, We note that the TRPMD and classical MD power spectra were obtained using a previous PES by Arandhara and Ramesh,\cite{Sai_2024} and those in the present calculations using our fit to their data. We verify that MULTIMODE results using the two PESs produce very similar results. This is shown in Table S6 in the ESI.

\subsection{AS-SCIVR Results}
\subsubsection{OH-Stretch}
\begin{figure}[H]
    \centering
    \includegraphics[width=\linewidth]{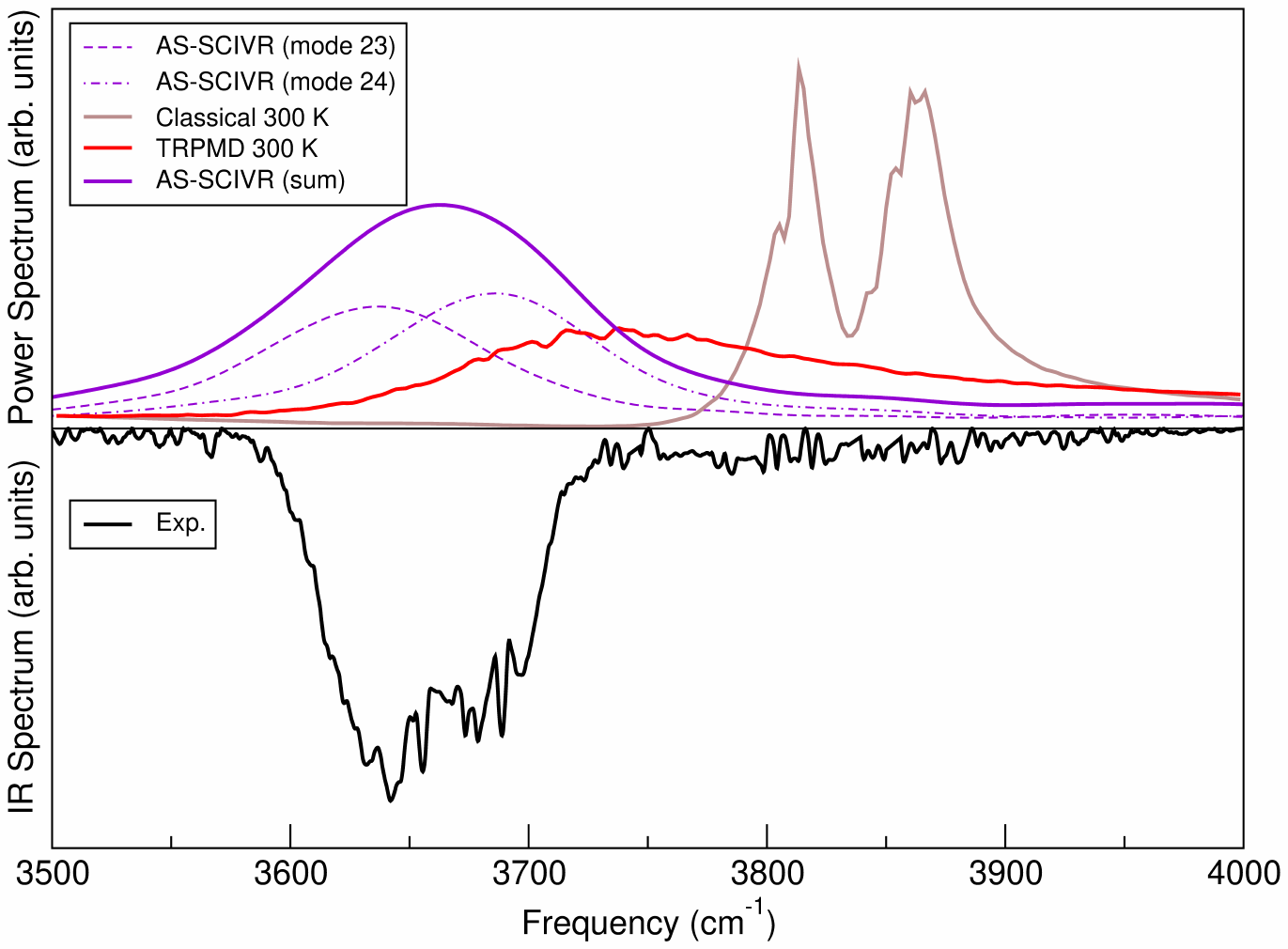}
    \caption{OH-stretch band for the \textbf{\ce{tG+g-}} conformer. On the top panel: AS-SCIVR results for mode 23 (violet, dash), mode 24 (violet, dash, and points), and their sum (violet, solid); TRPMD at 300K (red); classical (brown). On the bottom panel: experimental results (black).}
    \label{fig:AS_SCIVR_OH}
\end{figure}

Starting our description of the semiclassical results from the OH-stretch band, by looking at Figure \ref{fig:AS_SCIVR_OH} we notice that AS-SCIVR calculations for the global minimum (\textbf{\ce{tG+g-}}) conformer describe in an excellent way the experimental frequencies, differently from TRPMD and classical simulations which are sizeably shifted to larger frequencies. OH stretches are estimated by AS SCIVR at 3685 (mode 24) amd 3637 (mode 23) \cm, which are in excellent agreement with MULTIMODE values of 3681 and 3629 \cm, respectively.

To present a single curve also for the AS-SCIVR results we sum the two single-mode spectra and scale the outcome in a way that the area below it equals the area below the experimental curve in the 3500-4000 \cm~ range. This AS-SCIVR sum-of-states curve is represented with a solid line in Figure \ref{fig:AS_SCIVR_OH} with the calculations corresponding to the single modes reported in dashed and dash-and-points lines. The AS-SCIVR sum-of-states curve is a bit wider than the experimental one, but clearly narrower than the TRPMD one. In the case of the AS-SCIVR results, the increased width is due to the fact that the calculated power spectra include also all states close in frequency to the OH-stretch fundamentals which have a non-negligible projection onto the arbitrary quantum state $|\Psi(\bf{q}_{eq},\bf{p}_{eq})\rangle$ employed in Eq. \ref{eqn:astascivr}. Conversely, the IR experimental band is subject to selection rules and dipole strengths which decrease the number of states giving a non-negligible contribution to the band. These factors contribute to reduce the width of the experimental band.

\subsubsection{CH-Stretch}
\begin{figure}[H]
    \centering
    \includegraphics[width=\linewidth]{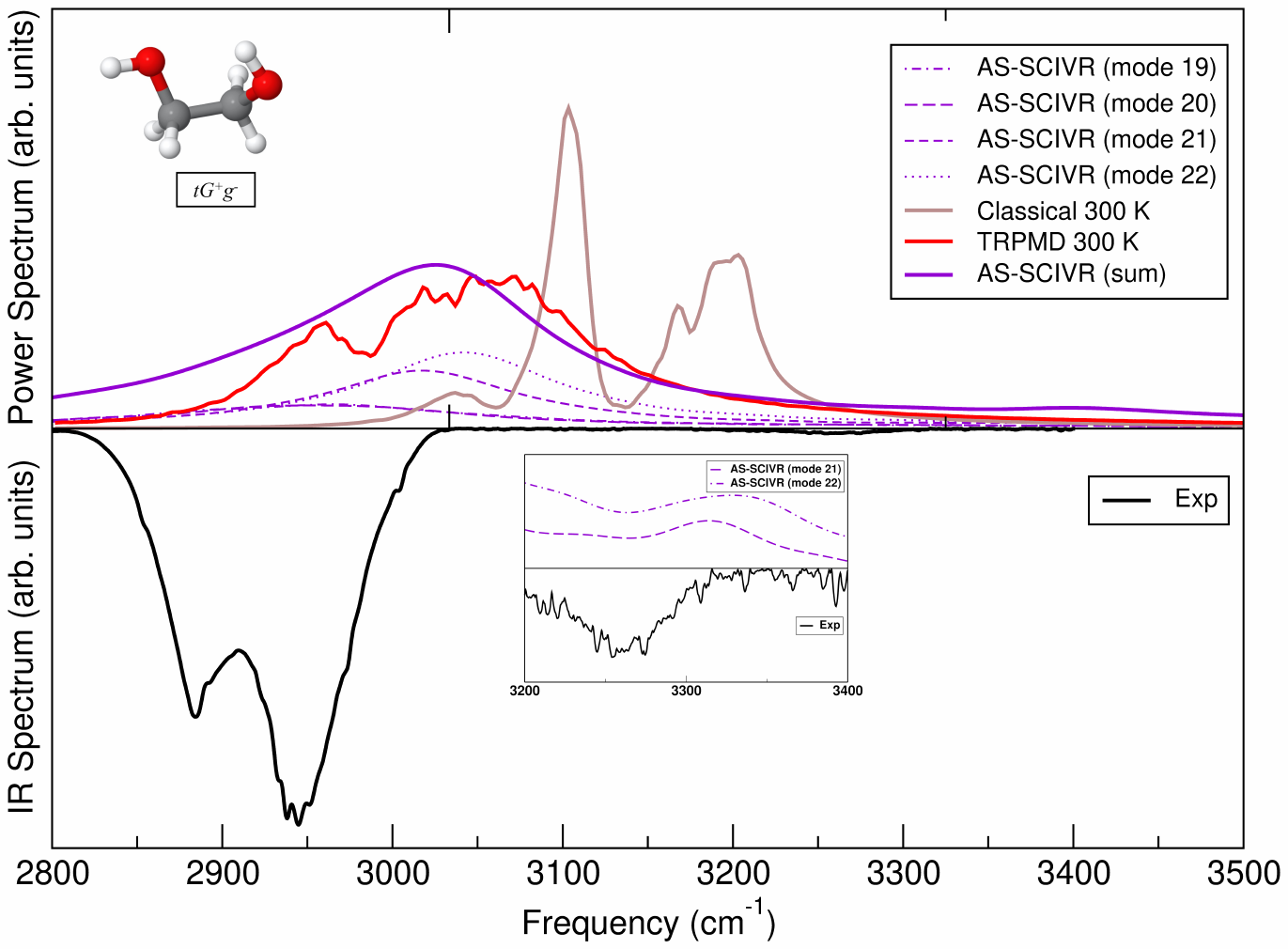}
    \caption{CH-stretch band. On top panel: AS-SCIVR results for modes 19-22 (violet, dashed and points) of the global minimum (\textbf{\ce{tG+g-}}), and their sum (violet, solid); TRPMD calculations at 300K (red); classical (brown). On bottom panel: experimental results (black); detail of the combination band in the 3200-3300 \cm region (inset plot).}
    \label{fig:AS_SCIVR_CH}
\end{figure}

Figure \ref{fig:AS_SCIVR_CH} refers to the CH-stretch band and it has been constructed in the same way of Figure \ref{fig:AS_SCIVR_OH}. AS-SCIVR calculations on the global \textbf{\ce{tG+g-}} minimum show that there are 4 fundamentals involved in the band. The four single-mode AS-SCIVR spectra can be separated into two groups with peak maxima shifted by about 60 \cm~ from the two maxima of the experimental spectrum. The sum-of-states spectrum presents again a single peak slightly more shifted from the experiment than Boltzmann-weighted smoothed MULTIMODE results, but slightly less shifted than TRPMD results. AS SCIVR estimates the fundamentals of modes 19-22 (the CH-stretch fundamentals) of the global minimum at 2931, 2941, 2989, and 3007 \cm, respectively. This is on average only 11 \cm~ different from MULTIMODE values.\\ AS-SCIVR results reported in Figure \ref{fig:AS_SCIVR_CH} refer to the global minimum conformer only. The necessity to look at other conformers to describe the lower frequency part of the CH-stretch band is confirmed by AS-SCIVR calculations (see below) as it was already pointed out by MULTIMODE ones. 

We notice that two fundamental spectral features, which are missed by TRPMD calculations, are remarkably found in the AS-SCIVR simulations. From the insight of Figure \ref{fig:AS_SCIVR_CH} it is evident that AS-SCIVR calculations for modes 21 and 22 present a combination band at about 3300 \cm. This corresponds to the experimental signal of low (but not negligible) intensity just below 3300 \cm. Remarkably, the shift in the AS-SCIVR estimate of this spectral feature is still 60 \cm as in the case of the CH-stretch fundamentals. The feature is interpreted as a combination band of these two modes with a low-frequency mode, arguably mode 3. This also explains why the combination is not found in MULTIMODE calculations, since they do not take into account modes with frequency below that of mode 10. Furthermore, no combination band is found in the TRPMD and classical results. 

The second feature of AS-SCIVR calculations we want to point out is related to Fermi resonances between the CH stretch and the bending overtone. We find, as shown in Figure S3 of the SI file, that besides the fundamental at 2941 \cm the AS-SCIVR simulation tailored for mode 19 of the global minimum conformer reports the fingerprint of Fermi resonances by showing two humps at 2884 and 2755 \cm. Likewise, MULTIMODE anticipates the involvement of mode 19 in Fermi resonances. Conversely, TRPMD is not able to reproduce this feature. Furthermore, the AS-SCIVR simulation detects also for mode 19 a combination band with a low-frequency mode, this time located at 3176 \cm.

\subsubsection{Other conformers}

Finally, we perform AS-SCIVR calculations on two other ethylene glycol conformers, namely the \textbf{\ce{g+Tg-}} and \textbf{\ce{tTt}} conformers. The goal is to find out if they could be responsible or at least contribute to the lower frequency peak in the experimental CH-stretch band, which is not described by the global minumum \textbf{\ce{tG+g-}} conformer. The two conformers lay at an electronic energy which is 1009 and 898 \cm higher than the global minimum, respectively. Calculations are more difficult because of a higher rejection rate of semiclassical trajectories in part due to the lower coverage of the PES for these two conformers. Thereby, we employ a semiclassical dynamics which is 20000 a.u. long rather than 25000 a.u. This allows us to improve the statistics (i.e. convergence) of our calculations at the cost of a lower, but still reasonable, spectral resolution. We find again the presence of Fermi resonances in the CH-stretch bands of these two conformers and, in addition, also stronger coupling between the CH modes. Figures S4 and S5 in the ESI file report these calculations. In particular, Figures S4 and S5 (the latter more clearly) demonstrate that AS-SCIVR calculations on these two conformers allow the regain also the lower-frequency part of the CH-stretch band.

For the \textbf{\ce{g+Tg-}} conformer we find the fundamental frequencies of modes 19-22 at 2929, 2957, 2999, and 3009 \cm. These values are very close to the MULTIMODE ones presented in Table S5 of the ESI file. However, these values appear to be still shifted from the lower peak of the experimental spectrum, which covers a range approximately between 2850 and 2910 \cm. Moving to the \textbf{\ce{tTt}} conformer, similar coupling features to those found for the previous conformer are present. The four fundamental frequencies are estimated by AS SCIVR to be at 2864, 2904, 2962, and 3000 \cm. Therefore, modes 19 and 20 appear to be suitable to describe, at least under the frequency aspect, the lower-frequency end of the experimental CH-stretch band. This is in good agreement with MULTIMODE calculations (see Table S4 in the ESI file). Differently from MULTIMODE calculations, though, in the AS-SCIVR calculations mode 19 and mode 20 besides being involved in the usual Fermi resonances appear to be sizably coupled also to mode 21. 

\section*{Summary and Conclusions}
We reported a permutationally invariant polynomial fit to 18,772 MP2/aug-cc-pVTZ energies and gradients for ethylene glycol.  This potential energy surface was used in VSCF/VCI and semi-classical AS-SCIVR calculations of the power spectrum in the spectral range of the CH and OH-stretches for low-lying conformers and compared to experiment and previous TRPMD and classical calculations of the power spectra.  The present calculations are in significantly better agreement with the experiment than these previous ones. While the OH-stretch band is dominated by a pure anharmonic OH-stretch, the CH-band is dominated by Fermi resonances with the overtone of bends. 

Regarding AS-SCIVR calculations we notice that they have been able to provide VCI-quality results, overperforming classical and TRPMD calculations. AS-SCIVR estimates have accurately described fundamental frequencies of vibrations for both the OH and CH-stretch bands, as well as Fermi resonances. Furthermore,  we remark that AS-SCIVR calculations were performed in full dimensionality, which is fundamental for the description of some spectroscopic features that may be missed by other methods. The latter include combination bands involving low-frequency motions and accurate estimates of the zero-point energy of each conformer. This work confirms the ability of semiclassical methods to accurately reproduce quantum effects when dealing with the spectroscopy (and also kinetics) of sizable molecules and chemical systems\cite{conte_Ceotto_perspectiveJPCL_2024,conte_Ceotto_perspectiveChemSci_2025} as in the present case for ethylene glycol or glycine in the past.\cite{conte_glycine20,mandelli_Aieta_glycinekinetics_2023}. It is also worth mentioning that recent progress in the semiclassical field has permitted to come up with an expression for the calculation of IR spectra,\cite{Lanzi_Conte_SCIR_2024} which is analogous to the one employed for power spectra. Therefore, application of the AS-SCIVR technique to IR calculations is anticipated in the near future.

Finally, the accuracy of both the VSCF/VCI and AS-SCIVR approaches exceeds that of previous TRPMD and classical MD ones for these bands. The origins of this difference in accuracy was described in detail. This finding is totally consistent with an earlier assessment for protonated water clusters, where however, only VSCF/VCI, TRPMD, and classical MD were compared.\cite{yutests}






\begin{acknowledgement}
A.N. thanks Prof. Alexandre Tkatchenko for the financial support from PHANTASTIC grant INTER/MERA22/16521502/PHANTASTIC.
J.M.B. thanks NASA, grant 80NSSC22K1167, for financial support. R.C. thanks Universit\`a degli Studi di Milano for funding under grant action PSR2023.

We thank Prof. Sai G. Ramesh and Mrinal Arandhara for providing us with the training dataset, their PES, and the data to plot the experimental spectrum as well as the classical and TRPMD power spectrum.
\end{acknowledgement}

\begin{suppinfo}
\begin{itemize}
\item Details of MULTIMODE input
\item Tables referred to in the text
\item Figures referred to in the text
\end{itemize}
\end{suppinfo}







\newpage
\section{Supporting Information}

\section{Details of the MULTIMODE Calculations}
Below are the essential input parameters for the MULTIMODE calculations.  More details are at https://scholarblogs.emory.edu/bowman/softwares/multimode/

NATOM,NSTAT,CONV,ICOUPL,ICOUPC,ISCFCI,IWHICH,IDISC,NROTTR,JMAX,INORM

    10    -1  1.D-3   4     2      250     1 
    
    0     -9     0     0      0
    
MAXBAS 

   10 10 10 10 10 10 10 10 10 10 10 10 10 10 10
   
   10 10 10 10 10 10 10 10 10 10 10 10 10 10 10
   
   10 10 10 10 10 10 10 10 10 10 10 10 10 10 10
   
   8 8 8 8 8 8 8 8 8 8 8 8 8 8 8
   
NBF,MBF,NVF

  12 18 6 12
  
  12 18 6 12
  
  12 18 6 12
  
  12 18 6 12
  
  12 18 6 12
  
  12 18 6 12
  
  12 18 6 12
  
  12 18 6 12
  
  12 18 6 12
  
  12 18 6 12
  
  12 18 6 12
  
  12 18 6 12
  
  12 18 6 12
  
  12 18 6 12
  
  12 18 6 12

\begin{table}[H]
\scriptsize
    \centering
    \begin{tabular}{ccccccccccc} 
    \toprule
    \toprule
    \textbf{Mode} &  \multicolumn{2}{c}{\textbf{tG+g-}} & \multicolumn{2}{c}{\textbf{g+G+g-}} & \multicolumn{2}{c}{\textbf{g-G+g-}} & \multicolumn{2}{c}{\textbf{tTt}} & \multicolumn{2}{c}{\textbf{g+Tg-}} \\
    \cmidrule(rl){2-3} \cmidrule(rl){4-5} \cmidrule(rl){6-7} \cmidrule(rl){8-9} \cmidrule(rl){10-11}
    & MP2/TZ$^a$ & PES & MP2/TZ$^a$ & PES & MP2/TZ$^a$ & PES & MP2/TZ$^a$ & PES & MP2/TZ$^a$ & PES \\
    \toprule
    1 & 168 & 170 & 168 & 166 & 100 & 87i & 116 & 133 & 141 & 145 \\
    2 & 247 & 214 & 292 & 304 & 159 & 166 & 217 & 183 & 250 & 190 \\
    3 & 329 & 333 & 327 & 328 & 321 & 327 & 230 & 207 & 268 & 215 \\
    4 & 420 & 404 & 452 & 464 & 428 & 426 & 291 & 296 & 296 & 287 \\
    5 & 523 & 531 & 536 & 532 & 528 & 532 & 481 & 483 & 475 & 470 \\
    6 & 887 & 890 & 878 & 874 & 881 & 879 & 839 & 843 & 803 & 789 \\
    7 & 904 & 904 & 897 & 901 & 885 & 890 & 1009 & 1046 & 1027 & 1043 \\
    8 & 1066 & 1072 & 1059 & 1063 & 1051 & 1059 & 1076 & 1075 & 1074 & 1075 \\
    9 & 1100 & 1099 & 1073 & 1074 & 1063 & 1073 & 1094 & 1091 & 1090 & 1083 \\
    10 & 1130 & 1136 & 1122 & 1122 & 1126 & 1123 & 1167 & 1169 & 1109 & 1107 \\
    11 & 1178 & 1211 & 1204 & 1217 & 1198 & 1204 & 1190 & 1237 & 1140 & 1157 \\
    12 & 1269 & 1278 & 1246 & 1253 & 1257 & 1283 & 1235 & 1243 & 1320 & 1337 \\
    13 & 1296 & 1315 & 1374 & 1362 & 1383 & 1401 & 1288 & 1308 & 1339 & 1343 \\
    14 & 1384 & 1392 & 1377 & 1382 & 1389 & 1402 & 1319 & 1329 & 1370 & 1367 \\
    15 & 1420 & 1430 & 1406 & 1408 & 1404 & 1415 & 1409 & 1429 & 1404 & 1382 \\
    16 & 1455 & 1467 & 1435 & 1440 & 1422 & 1432 & 1487 & 1513 & 1433 & 1432 \\
    17 & 1516 & 1520 & 1511 & 1516 & 1508 & 1513 & 1541 & 1546 & 1521 & 1511 \\
    18 & 1524 & 1526 & 1521 & 1522 & 1512 & 1527 & 1551 & 1551 & 1535 & 1537 \\
    19 & 3053 & 3056 & 3026 & 3027 & 3071 & 3069 & 3057 & 3065 & 3067 & 3068 \\
    20 & 3058 & 3064 & 3070 & 3074 & 3074 & 3078 & 3064 & 3066 & 3076 & 3076 \\
    21 & 3114 & 3118 & 3134 & 3137 & 3140 & 3141 & 3102 & 3105 & 3124 & 3127 \\
    22 & 3149 & 3149 & 3159 & 3161 & 3148 & 3151 & 3127 & 3130 & 3150 & 3150 \\
    23 & 3808 & 3835 & 3794 & 3817 & 3845 & 3856 & 3857 & 3872 & 3840 & 3859 \\
    24 & 3856 & 3872 & 3831 & 3849 & 3846 & 3882 & 3858 & 3879 & 3842 & 3870 \\
    \toprule
    \bf{MEA} & & \bf{10} & & \bf{6} & & \bf{9} & & \bf{13} & & \bf{12} \\
    \toprule
    \end{tabular}
    \\
    $^a$ From Table S-3 in Ref. \citenum{Sai_2024}\\
    \caption{Normal mode frequencies (cm$^{-1}$) for five low-lying conformers of ethylene glycol from unweighted fitting PES.}
    \label{tab:Freq}
\end{table}

\newpage
\section{VCI frequencies}

\begin{table}[H]
    \centering
    \begin{tabular}{ccccc} 
    \toprule
    \toprule
    \textbf{Mode} & \textbf{Har. Freq.} & \textbf{CI Freq.} & \textbf{CI Coeff.} & \textbf{Coupling Modes} \\
    \toprule
    10 & 1121 & 1115 &  0.9923 & $\nu_{10}$ \\
    11 & 1215 & 1177 & -0.9829, -0.0936, -0.0695  & $\nu_{11}$, $\nu_{12}$, $\nu_{14}$ \\
    12 & 1250 & 1214 &  0.9823, -0.1022, -0.0752  & $\nu_{12}$, $\nu_{11}$, $\nu_{16}$ \\
    13 & 1360 & 1315 & -0.9765, -0.1711,  0.0833  & $\nu_{13}$, $\nu_{14}$, $\nu_{11}$ \\
    14 & 1379 & 1335 &  0.9240, -0.2688, -0.1694  & $\nu_{14}$, $\nu_{15}$, $\nu_{13}$ \\
    15 & 1405 & 1372 &  0.9304,  0.2980,  0.1838  & $\nu_{15}$, $\nu_{14}$, $\nu_{16}$ \\
    16 & 1437 & 1405 &  0.9611, -0.2273,  0.1066 & $\nu_{16}$, $\nu_{15}$, $\nu_{14}$ \\
    17 & 1514 & 1470 &  0.9961 & $\nu_{17}$ \\
    18 & 1519 & 1472 &  0.9967 & $\nu_{18}$ \\
    \\
    \multirow{2}{*}{19} & \multirow{2}{*}{3027} & 2894 &  0.6762, 0.4132, 0.3474 & $\nu_{19}$, ($\nu_{16}$ + $\nu_{18}$), 2$\nu_{18}$ \\
    & & 2943 & 0.8374, -0.3496, 0.3443 & 2$\nu_{18}$, $\nu_{19}$, $\nu_{21}$ \\
    \\
    \multirow{2}{*}{20} & \multirow{2}{*}{3073} & 2910 & 0.6971, -0.5907, -0.2296 & 2$\nu_{17}$, $\nu_{20}$, ($\nu_{17}$ + $\nu_{18}$),  \\
     & & 2971 & 0.6708, 0.4928, -0.4029 & $\nu_{20}$, 2$\nu_{17}$, ($\nu_{17}$ + $\nu_{18}$) \\
     \\
     \multirow{2}{*}{21} & \multirow{2}{*}{3136} & 2943 & 0.8374, -0.3496, 0.3443 & 2$\nu_{18}$, $\nu_{19}$, $\nu_{21}$ \\
      & & 3003 & -0.8705, 0.3152, -0.1917 & $\nu_{21}$, 2$\nu_{18}$, ($\nu_{17}$ + $\nu_{18}$) \\
      \\
      22 & 3160 & 3017 & 0.9119, -0.1847, -0.1522 & $\nu_{22}, 2\nu_{17}, \nu_{20}$ \\
      23 & 3815 & 3614 & 0.9714, 0.1338, 0.0950 & $\nu_{23}, (\nu_{11} + 2\nu_{12}), \nu_{24}$ \\
      24 & 3847 & 3655 &  0.9385, -0.2650, -0.0923 & $\nu_{24}, (2\nu_{11} + \nu_{13}), \nu_{23}$ \\
    \toprule
    \end{tabular}
    \caption{VCI Frequency (cm$^{-1}$) and leading coefficients for \textbf{\ce{g+G+g-}} conformer.}
    \label{tab:CI_Freq2}
\end{table}

\begin{table}[H]
    \centering
    \begin{tabular}{ccccc} 
    \toprule
    \toprule
    \textbf{Mode} & \textbf{Har. Freq.} & \textbf{CI Freq.} & \textbf{CI Coeff.} & \textbf{Coupling Modes} \\
    \toprule
    10 & 1123 & 1117 & -0.9956 & $\nu_{10}$ \\
    11 & 1200 & 1168 &  0.9917 & $\nu_{11}$ \\
    12 & 1280 & 1237 & -0.9818 & $\nu_{12}$ \\
    13 & 1401 & 1371 &  0.9564, -0.2311, -0.1472 & $\nu_{13}$, $\nu_{14}$, $\nu_{15}$ \\
    14 & 1403 & 1351 & -0.8562, -0.3324, 0.2763  & $\nu_{14}$, $\nu_{15}$, $\nu_{16}$ \\
    15 & 1414 & 1353 & -0.8889, -0.3144, 0.2662  & $\nu_{15}$, $\nu_{16}$, $\nu_{14}$ \\
    16 & 1428 & 1394 &  0.9004, 0.3561, -0.2146  & $\nu_{16}$, $\nu_{14}$, $\nu_{15}$ \\
    17 & 1513 & 1467 &  0.9897 & $\nu_{17}$ \\
    18 & 1526 & 1477 & -0.9887 & $\nu_{18}$ \\
    \\
    \multirow{2}{*}{19} & \multirow{2}{*}{3069} & 2908 & 0.6926, -0.5415, 0.2234 & 2$\nu_{17}$, $\nu_{19}$, ($\nu_{17}$ + $\nu_{18}$) \\
    & & 2964 & 0.5759, -0.5147, 0.3646 & $\nu_{19}$, 2$\nu_{18}$, ($\nu_{17}$ + $\nu_{18}$) \\
    \\
    \multirow{2}{*}{20} & \multirow{2}{*}{3077} & 2920 & 0.6236, -0.5043, -0.4766 & $\nu_{20}$, ($\nu_{17}$ + $\nu_{18}$), 2$\nu_{18}$  \\
     & & 2983 & 0.6015, 0.5661, 0.3465 & $\nu_{20}$, ($\nu_{17}$ + $\nu_{18}$), 2$\nu_{18}$ \\
     \\
     \multirow{2}{*}{21} & \multirow{2}{*}{3142} & 3001 & -0.8345, 0.2630, -0.2177 & $\nu_{21}$, $\nu_{22}$, 2$\nu_{18}$ \\
      & & 3009 & 0.8584, 0.2584, -0.1771 & $\nu_{22}$, $\nu_{21}$, $\nu_{19}$  \\
      \\
      \multirow{2}{*}{22} & \multirow{2}{*}{3150} & 3001 & -0.8345, 0.2630, -0.2177 & $\nu_{21}$, $\nu_{22}$, 2$\nu_{18}$ \\
      & & 3009 & 0.8584, 0.2584, -0.1771 & $\nu_{22}$, $\nu_{21}$, $\nu_{19}$  \\
      \\
      \multirow{3}{*}{23} & \multirow{3}{*}{3856} & 3657 & 0.6533, -0.4486, -0.2863 & 
      $(\nu_{10} + \nu_{11} + \nu_{15}), \nu_{23}, (2\nu_{11} + \nu_{15})$ \\
       & & 3660 & 0.7299, 0.5507, -0.1638 & $\nu_{23}, (\nu_{10} + \nu_{11} + \nu_{15}),\nu_{24}$ \\
       & & 3669 & 0.6145, -0.4001, 0.2924 & $(2\nu_{11} + \nu_{15}), \nu_{23}, (2\nu_{11} + \nu_{16})$ \\
       \\
      \multirow{2}{*}{24} & \multirow{2}{*}{3876} & 3681 & -0.7377, -0.3362, 0.2682 & $3\nu_{12}, \nu_{24}, (2\nu_{12} + \nu_{15})$ \\
      & & 3683 & 0.8559, -0.3766, -0.1531 & $\nu_{24}, 3\nu_{12}, (2\nu_{11} + \nu_{14})$ \\
    \toprule
    \end{tabular}
    \caption{VCI Frequency (cm$^{-1}$) for \textbf{\ce{g- G+g-}} conformer.}
    \label{tab:CI_Freq3}
\end{table}

\begin{table}[H]
    \centering
    \begin{tabular}{ccccc} 
    \toprule
    \toprule
    \textbf{Mode} & \textbf{Har. Freq.} & \textbf{CI Freq.} & \textbf{CI Coeff.} & \textbf{Coupling Modes} \\
    \toprule
    10 & 1168 & 1152 &  0.9943 & $\nu_{10}$ \\
    11 & 1236 & 1183 &  0.9857 & $\nu_{11}$ \\
    12 & 1241 & 1218 &  0.9941 & $\nu_{12}$ \\
    13 & 1306 & 1258 &  0.9863 & $\nu_{13}$ \\
    14 & 1328 & 1296 & -0.9944 & $\nu_{14}$ \\
    15 & 1427 &	1387 & -0.9884 & $\nu_{15}$ \\
    16 & 1513 & 1473 & -0.9782 & $\nu_{16}$ \\
    17 & 1545 & 1503 & -0.9788 & $\nu_{17}$ \\
    18 & 1549 & 1506 & -0.9905 & $\nu_{18}$ \\
    \\
    \multirow{3}{*}{19} & \multirow{3}{*}{3065} & 2894 & -0.7185, -0.5062, -0.2092 & $\nu_{19}, 2\nu_{16}, 2\nu_{17}$ \\
    & & 2961 & 0.8046, -0.4110, -0.2573 & $2\nu_{16}, \nu_{19}, 2\nu_{17}$ \\
    & & 3016 & -0.6320, -0.6191, 0.3689 & $2\nu_{17}, 2\nu_{18}, \nu_{19}$\\
    \\
    \multirow{3}{*}{20} & \multirow{3}{*}{3066} & 2842 & 0.7807, 0.5193, 0.2552 & $(\nu_{15} + \nu_{16}), \nu_{20}, (\nu_{15} + \nu_{17})$ \\
    & & 2942 & 0.7045, -0.4065, -0.3483 & $\nu_{20}, (\nu_{15} + \nu_{16}), (\nu_{15} + \nu_{17})$\\
    & & 3034 & 0.9191, -0.2876, -0.1614 & $(\nu_{17} + \nu_{18}), \nu_{20}, (\nu_{16} + \nu_{18})$ \\
    \\
    21 & 3104 & 2968 & -0.8986, -0.2165, -0.1774 & $ \nu_{21}, (\nu_{14} + \nu_{17}), (\nu_{14} + \nu_{16})$ \\
    22 & 3130 & 2990 &  0.9218, -0.1838, -0.1759 & $ \nu_{22}, (\nu_{12} + \nu_{16}), (\nu_{10} + \nu_{18})$ \\
    23 & 3870 & 3687 & -0.8991, 0.3343, 0.1510 & $ \nu_{23}, \nu_{24}, (\nu_{23} + \nu_{24})$ \\
    24 & 3878 & 3690 & 0.9157, 0.3283, -0.1073 & $ \nu_{24}, \nu_{23}, 2\nu_{23} $ \\
    \toprule
    \end{tabular}
    \caption{CI Frequency (cm$^{-1}$) for \textbf{\ce{tTt}} conformer.}
    \label{tab:CI_Freq4}
\end{table}

\begin{table}[H]
    \centering
    \begin{tabular}{ccccc} 
    \toprule
    \toprule
    \textbf{Mode} & \textbf{Har. Freq.} & \textbf{CI Freq.} & \textbf{CI Coeff.} & \textbf{Coupling Modes} \\
    \toprule
    10 & 1107 & 1106 & -0.9899 & $\nu_{10}$ \\
    11 & 1154 & 1125 &  0.9856 & $\nu_{11}$ \\
    12 & 1337 &	1296 & -0.7891, 0.5335, -0.2666 & $ \nu_{12}, \nu_{14}, \nu_{13} $ \\
    13 & 1343 & 1312 & -0.9336, -0.2421, 0.1883 & $ \nu_{13}, \nu_{15}, \nu_{12} $ \\
    14 & 1364 & 1317 &  0.8238, 0.5424, -0.1183 & $ \nu_{14}, \nu_{12}, \nu_{15} $ \\
    15 & 1381 & 1341 & -0.9503, 0.2143, -0.1889 & $ \nu_{15}, \nu_{13}, \nu_{12} $ \\
    16 & 1429 & 1405 & -0.9934 & $ \nu_{16} $ \\
    17 & 1507 & 1470 & -0.9943 & $ \nu_{17} $ \\
    18 & 1535 & 1492 &  0.9946 & $ \nu_{18} $ \\
    \\
    \multirow{4}{*}{19} & \multirow{4}{*}{3067} & 2790 & -0.9130, -0.3094, 0.0815 & $ 2\nu_{16}, \nu_{19}, (\nu_{12} + \nu_{18}) $ \\
    & & 2912 & -0.6502, -0.6075, 0.2754 & $ 2\nu_{17}, \nu_{19}, 2\nu_{16} $ \\
    & & 2954 & 0.7031, -0.5006, -0.3991 & $ 2\nu_{17}, \nu_{19}, 2\nu_{18} $ \\
    & & 2993 & -0.7501, -0.4068, 0.4004 & $ 2\nu_{18}, \nu_{21}, \nu_{19} $ \\
    \\
    \multirow{3}{*}{20} & \multirow{3}{*}{3075} & 2910 & -0.9348, 0.2284, 0.1855 & $ (\nu_{16} + \nu_{18}), \nu_{20}, (\nu_{17} + \nu_{18}) $ \\
    & & 2925 & -0.7167, -0.5180, -0.3171 & $ \nu_{20}, (\nu_{17} + \nu_{18}), (\nu_{16} + \nu_{18}) $ \\
    & & 2996 & -0.6766, 0.5223, 0.4347 & $ (\nu_{17} + \nu_{18}), \nu_{22}, \nu_{20} $ \\
    \\
    \multirow{2}{*}{21} & \multirow{2}{*}{3127} & 2980 & -0.8179, 0.4412, -0.1271 & $ \nu_{21}, 2\nu_{18}, \nu_{19} $ \\
    & & 2993 & -0.7501, -0.4068, 0.4004 & $ 2\nu_{18}, \nu_{21}, \nu_{19} $ \\
    \\
    \multirow{2}{*}{22} & \multirow{2}{*}{3149} & 2996 & -0.6766, 0.5223, 0.4347 & $ (\nu_{17} + \nu_{18}), \nu_{22}, \nu_{20} $ \\
    & & 3007 & -0.7698, -0.4640, 0.2974 & $ \nu_{22}, (\nu_{17} + \nu_{18}), \nu_{19} $ \\
    \\
    23 & 3856 & 3666 & -0.9619 & $\nu_{23}$ \\
    24 & 3869 & 3675 & -0.9734 & $\nu_{24}$ \\
    \toprule
    \end{tabular}
    \caption{CI Frequency (cm$^{-1}$) for \textbf{\ce{g+Tg-}} conformer.}
    \label{tab:CI_Freq5}
\end{table}

\begin{table}[H]
    \centering
    \begin{tabular}{cccc}
        \toprule
        \toprule
         \textbf{Mode} & \textbf{Har. Freq.} & \multicolumn{2}{c}{\textbf{VCI Freq.}} \\
         \cmidrule(rl){3-4}
          &   & \textbf{PIP PES} & \textbf{Sai PES\cite{Sai_2024}} \\ \toprule
        10 & 1135 & 1119 & 1118 \\ 
        11 & 1207 & 1167 & 1155 \\ 
        12 & 1277 & 1246 & 1249 \\ 
        13 & 1314 & 1268 & 1271 \\ 
        14 & 1390 & 1340 & 1355 \\ 
        15 & 1429 & 1387 & 1392 \\ 
        16 & 1465 & 1429 & 1435 \\ 
        17 & 1519 & 1472 & 1477 \\ 
        18 & 1525 & 1478 & 1485 \\ 
           &      &      &      \\
        \multirow{4}{*}{19} & \multirow{4}{*}{3056} & 2798 & 2810 \\ 
         &  & 2918 & 2917 \\ 
         &  & 2972 & 2923 \\
         &  & 2981 & 2986 \\ 
            &      &      &      \\
        \multirow{4}{*}{20} & \multirow{4}{*}{3064} & 2836 & 2849 \\ 
         &  & 2915 & 2912 \\ 
         &  & 2961 & 2958 \\ 
         &  &  &  \\ 
        \multirow{4}{*}{21} & \multirow{4}{*}{3118} & 2972 & 2958 \\
         &  & 2981 & 2967 \\
         &  &  & 2973 \\
         &  &  &  \\ 
        22 & 3148 & 3012 & 3013 \\ 
        23 & 3831 & 3629 & 3623 \\ 
        24 & 3871 & 3681 & 3674 \\
        \toprule
    \end{tabular}
    \caption{VCI frequency (cm$^{-1}$) comparison between PIP PES and Sai's PES\cite{Sai_2024} for conformer \textbf{\ce{tG+g-}}.}
    \label{tab:freq_two_pes}
\end{table}

\begin{table}[H]
    \centering
    \begin{tabular}{ccccc} 
    \toprule
    \toprule
    \textbf{Mode} & \textbf{Har. Freq.} & \textbf{CI Freq.} & \textbf{VCI Coeff.} & \textbf{Corresponding Modes} \\
    \toprule
    10 & 1135 & 1119 & 0.9957 & $\nu_{10}$ \\
    11 & 1207 & 1166 & -0.9889, 0.0582, 0.0455  & $\nu_{11}$, $\nu_{12}$, $\nu_{14}$ \\
    12 & 1277 & 1245 &  0.9775, 0.1739, 0.0633  & $\nu_{12}$, $\nu_{13}$, $\nu_{11}$ \\
    13 & 1314 & 1268 & 0.9633, -0.1765, -0.1319  & $\nu_{13}$, $\nu_{12}$, $\nu_{15}$ \\
    14 & 1390 & 1339 &  -0.9345, -0.3040, -0.1419  & $\nu_{14}$, $\nu_{15}$, $\nu_{13}$ \\
    15 & 1429 & 1387 & 0.9355, -0.3206, 0.0953 & $\nu_{15}$, $\nu_{14}$, $\nu_{13}$ \\
    16 & 1465 & 1429 & -0.9913, -0.0479, 0.0418 & $\nu_{16}$, $\nu_{18}$, $\nu_{15}$ \\
    17 & 1519 & 1471 &  0.9549, -0.2797, 0.0396 & $\nu_{17}$, $\nu_{18}$, $\nu_{15}$ \\
    18 & 1525 & 1477 &  0.9546, 0.2807, -0.0398 & $\nu_{18}$, $\nu_{17}$, $\nu_{16}$ \\
    \\
    \multirow{3}{*}{19} & \multirow{3}{*}{3056} & 2797 &  -0.8310, -0.3904, -0.2719 & ($\nu_{15}$ + $\nu_{16}$), $\nu_{19}$, 2$\nu_{16}$ \\
    & & 2917 & -0.5982,  0.4581, 0.3712 & $\nu_{19}$, 2$\nu_{18}$, 2$\nu_{16}$ \\
    & & 2971 & 0.6708, 0.3955, 0.3528 & $\nu_{21}$, 2$\nu_{18}$, $\nu_{19}$ \\
    & & 2980 & 0.5377, -0.5366, -0.4214 & $\nu_{21}$, ($\nu_{17}$ + $\nu_{18}$), $\nu_{19}$ \\
    \\
    \multirow{2}{*}{20} & \multirow{2}{*}{3064} & 2836 & 0.8094, -0.3438, -0.2594 & 2$\nu_{16}$, ($\nu_{15}$ + $\nu_{16}$), $\nu_{20}$ \\
     & & 2905 & -0.8229, 0.2845, -0.2218 & ($\nu_{16}$ + $\nu_{17}$), $\nu_{20}$, ($\nu_{17}$ + $\nu_{18}$) \\
     & & 2914 & -0.5085, 0.5010, -0.4777 & ($\nu_{16}$ + $\nu_{17}$), 2$\nu_{17}$, $\nu_{20}$ \\
     & & 2960 & 0.5301, 0.4905, -0.3889 & $\nu_{20}$, ($\nu_{17}$ + $\nu_{18}$), 2$\nu_{18}$ \\
     \\
     \multirow{2}{*}{21} & \multirow{2}{*}{3118} & 2971 & 0.6708, 0.3955, 0.3528 & $\nu_{21}$, 2$\nu_{18}$, $\nu_{19}$ \\
      & & 2980 & 0.5377, -0.5366, -0.4214 & $\nu_{21}$, ($\nu_{17}$ + $\nu_{18}$), $\nu_{19}$ \\
      \\
      22 & 3148 & 3011 & 0.8886, -0.1972,  -0.1733 & $\nu_{22}, 2\nu_{17}, (\nu_{17} + \nu_{18})$ \\
      23 & 3831 & 3629 & -0.8159, -0.4973, 0.1774 & $\nu_{23}, (2\nu_{10} + \nu_{15}), (2\nu_{10} + \nu_{14})$ \\
      24 & 3871 & 3681 &  -0.9622, -0.1575,  0.0869 & $\nu_{24}, (\nu_{11} + 2\nu_{13}), \nu_{23}$ \\
    \toprule
    \end{tabular}
    \caption{VSCF/VCI Energies (cm$^{-1}$) and VCI expansion coefficients for \textbf{\ce{tG+g-}} conformer without Coriolis coupling.}
    \label{tab:CI_Freq1}
\end{table}

\newpage
\section{MULTIMODE Spectra}

\begin{figure}[H]
    \centering
    \includegraphics[width=1.0\linewidth]{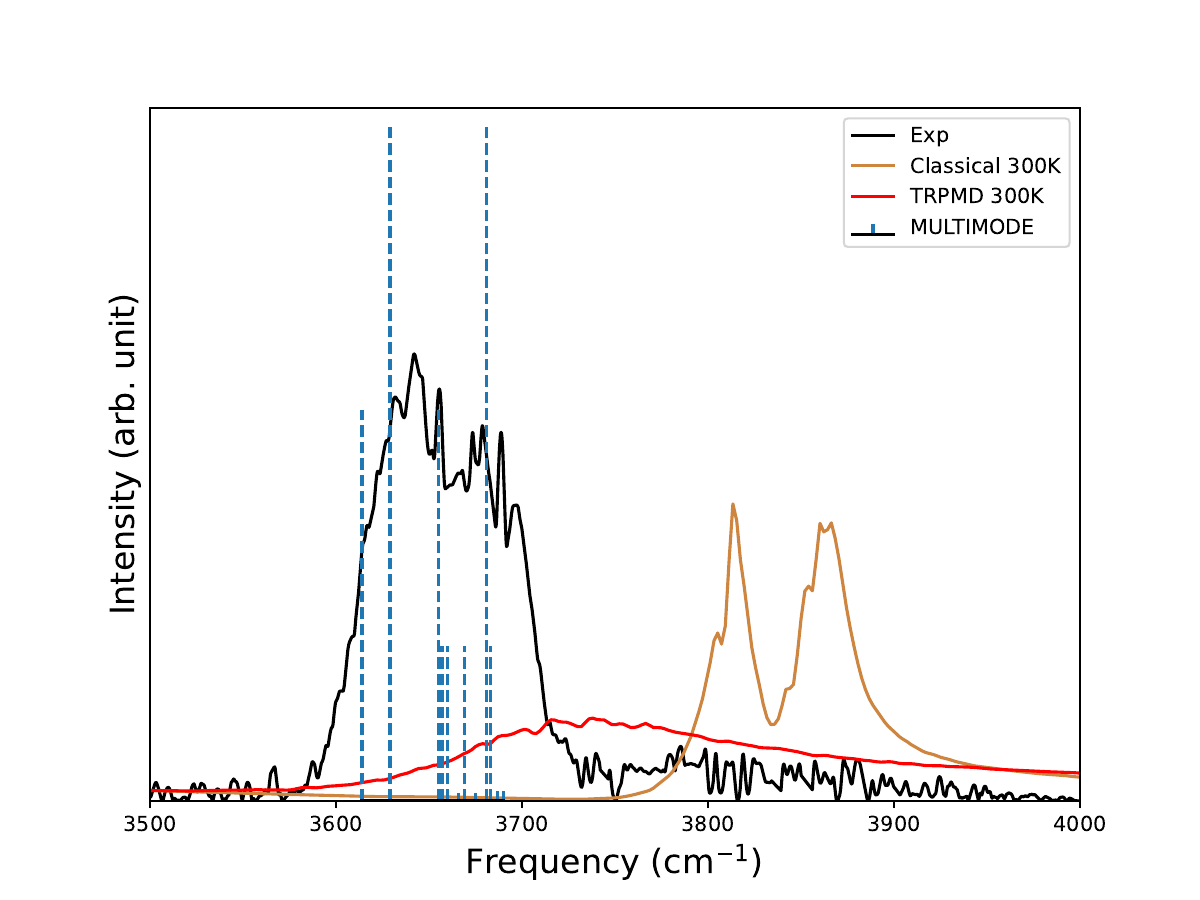}
    \caption{MULTIMODE eigenstates (blue dotted sticks) computed from VSCF/VCI calculations, power spectra at 300 K from Classical MD (orange curve) and TRPMD (red curve) simulations from Arandhara \textit{et al}\cite{Sai_2024}compared with experimental IR spectra (black curve) from Das \textit{et al}\cite{das2015} for \ce{O-H} stretching region.}
    \label{fig:freq_OH_1}
\end{figure}

\begin{figure}[H]
    \centering
    \includegraphics[width=1.0\linewidth]{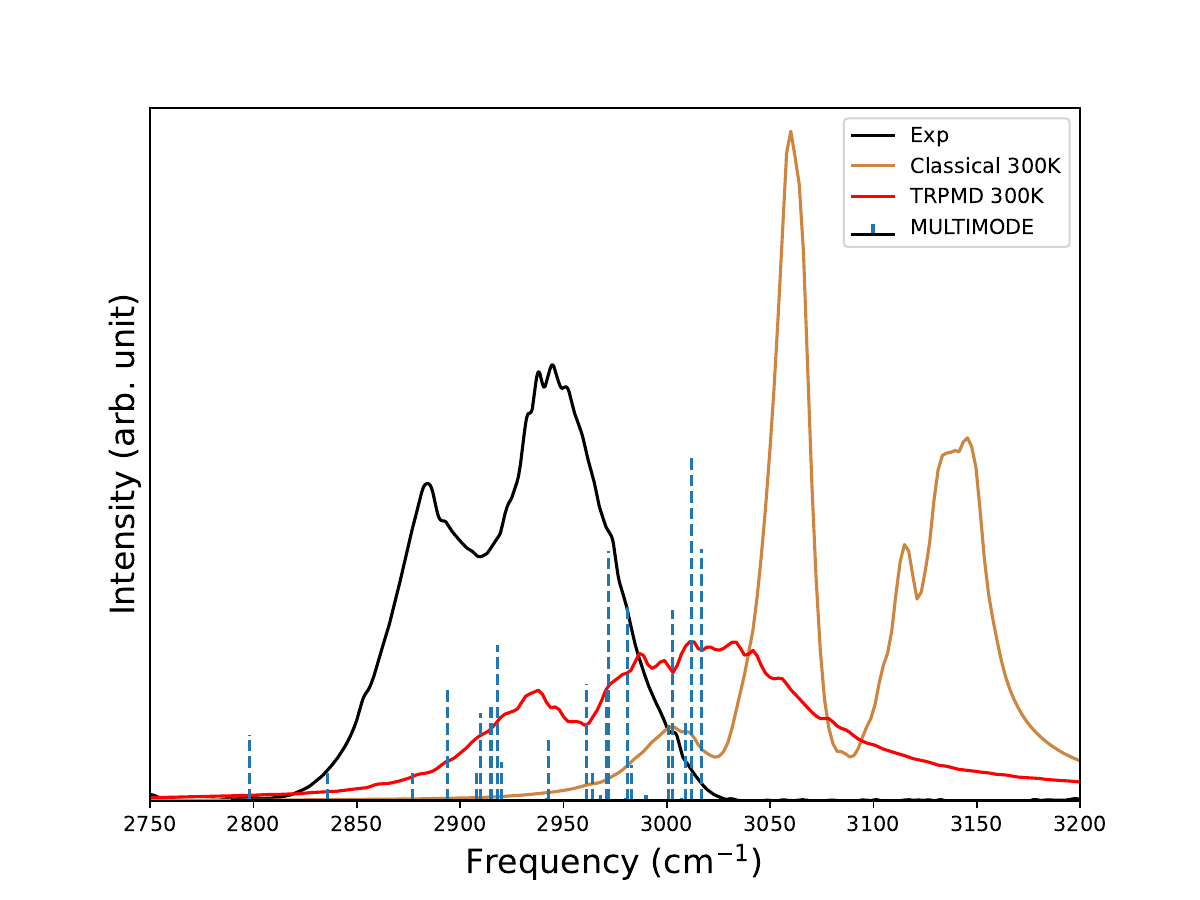}
    \caption{MULTIMODE eigenstates (blue dotted sticks) computed from VSCF/VCI calculations, power spectra at 300 K from Classical MD (orange curve) and TRPMD (red curve) simulations from Arandhara \textit{et al}\cite{Sai_2024}compared with experimental IR spectra (black curve) from Das \textit{et al}\cite{das2015} for \ce{C-H} stretching region.}
    \label{fig:freq_CH_1}
\end{figure}



\newpage
\section{AS-SCIVR CH-stretch band Fermi resonances}

The following Figure shows that AS-SCIVR calculations performed on the global minimum \textbf{\ce{tG+g-}} conformer are able to identify Fermi resonances at 2755 \cm and 2884 \cm , as well as a combination band of mode 19 with a low-frequency mode (arguably mode 3) found at 3176 \cm.
\begin{figure}[H]
    \centering
    \includegraphics[width=1.0\linewidth]{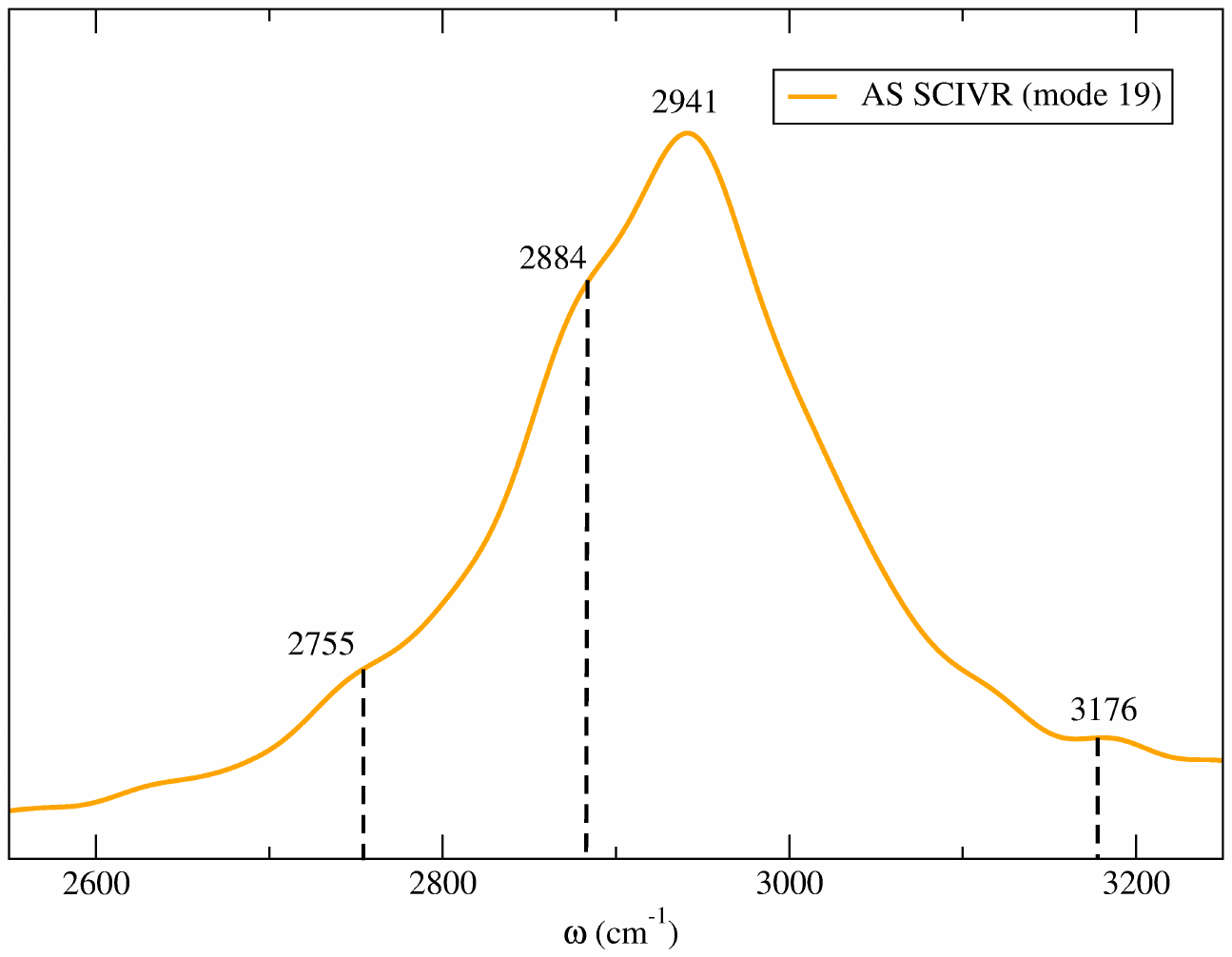}
    \caption{Fundamental frequency, Fermi resonances, and combination band involving mode 19 of the global minimum \textbf{\ce{tG+g-}} conformer. }
    \label{fig:AS_SCIVR_Fermimode19}
\end{figure}
\newpage
\section{AS-SCIVR CH-stretch band: \ce{g+Tg-} Conformer }
The following Figure for the CH-stretch band shows a comparison between different power spectrum calculations and the experiment. AS-SCIVR calculations have been performed on modes 19-22 of the \ce{g+Tg-} conformer. 
\begin{figure}[H]
    \centering
    \includegraphics[width=1.1\linewidth]{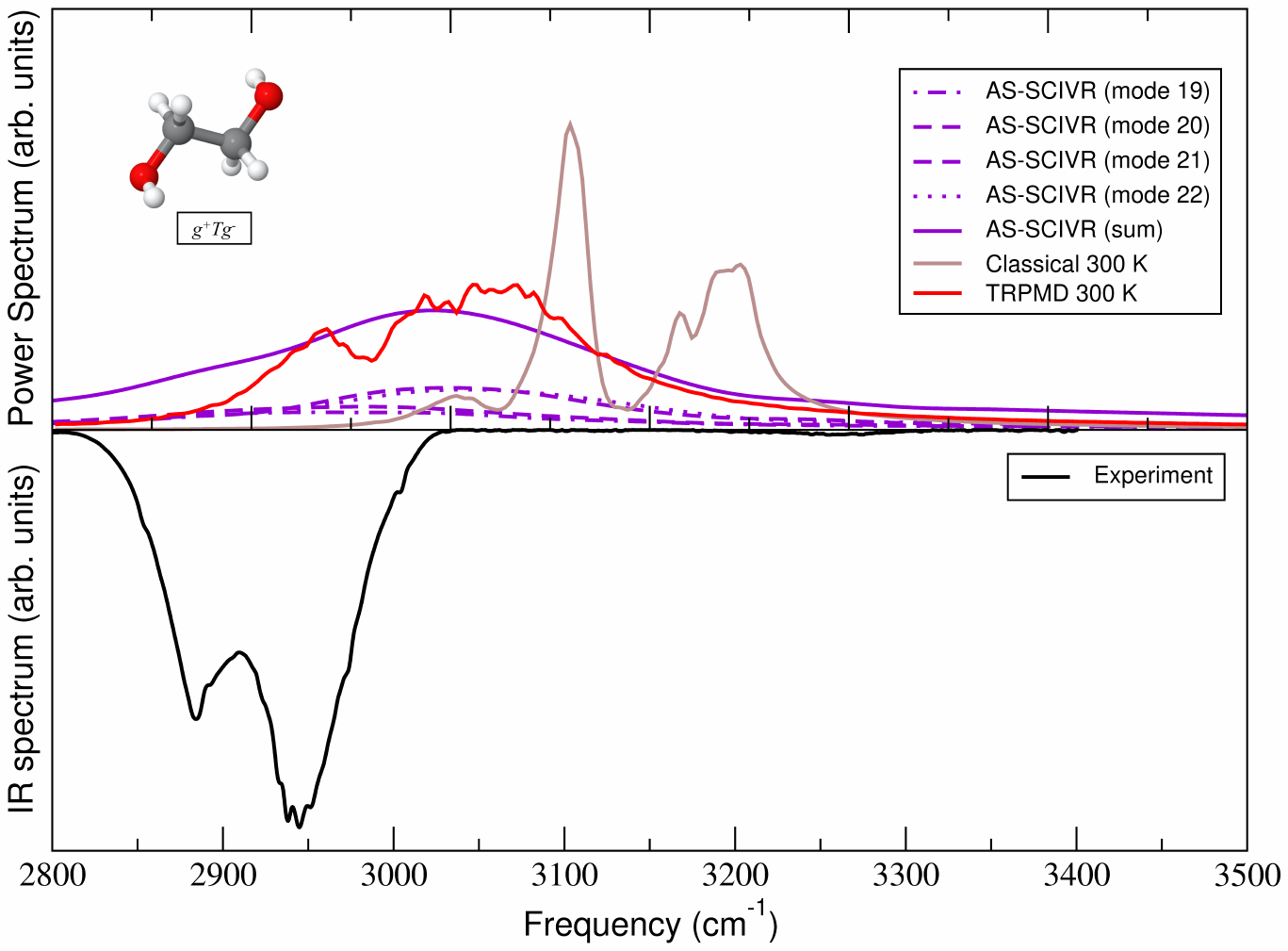}
    \caption{CH-stretch band. On top panel: AS-SCIVR results for modes 19-22 (violet, dashed and points) of the \textbf{\ce{g+Tg-}} conformer, and their sum (violet, solid); TRPMD calculations at 300K (red); classical (brown). On bottom panel: experimental results (black).}
    \label{fig:AS_SCIVR_g+Tg-}
\end{figure}
\newpage
\section{AS-SCIVR CH-stretch band: \ce {tTt} conformer}
The following Figure for the CH-stretch band shows a comparison between different power spectrum calculations and the experiment. AS-SCIVR calculations have been performed on modes 19-22 of the \ce{tTt} conformer. 
\begin{figure}[H]
    \centering
    \includegraphics[width=1.1\linewidth]{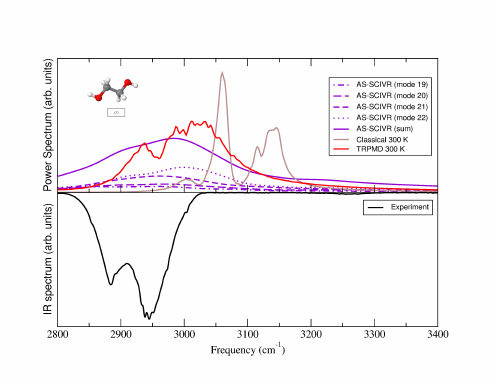}
    \caption{CH-stretch band. On top panel: AS-SCIVR results for modes 19-22 (violet, dashed and points) of the \textbf{\ce{tTt}} conformer, and their sum (violet, solid); TRPMD calculations at 300K (red); classical (brown). On bottom panel: experimental results (black).}
    \label{fig:AS_SCIVR_tTt}
\end{figure}

\bibliography{ref_tot}

\end{document}